\documentclass[useAMS,usenatbib]{mn2e}

\input{epsf}

\usepackage{natbib}
\usepackage{graphicx}	
\usepackage{deluxetable}
\usepackage{amsmath}
\usepackage{longtable}
\usepackage{lscape}
\usepackage{threeparttable}
\usepackage{threeparttablex}
\usepackage{deluxetable}
\usepackage[singlelinecheck=false]{caption}
\usepackage{afterpage}

\def\lsim{\lower0.3em\hbox{$\,\buildrel <\over\sim\,$}}
\def\gsim{\lower0.3em\hbox{$\,\buildrel >\over\sim\,$}}

\title[The orientation dependence of quasar SEDs]{The orientation dependence of quasar spectral energy distributions}
\author[J. C. Runnoe et al.]{Jessie C. Runnoe$^{1}$\thanks{E-mail:
jrunnoe@uwyo.edu} , Z. Shang$^{2}$, and M. S. Brotherton$^{1}$\\
$^{1}$Department of Physics and Astronomy, University of Wyoming, Laramie, WY 82071, USA\\
$^{2}$Department of Physics, Tianjin Normal University, Tianjin 300387, China}
\begin{document}		

\date{Preprint 2013 June 10}

\pagerange{\pageref{firstpage}--\pageref{lastpage}} \pubyear{2012}

\maketitle

\label{firstpage}

\begin{abstract}
We investigate the orientation dependence of the spectral energy distributions in a sample of radio-loud quasars.  Selected specifically to study orientation issues, the sample contains 52 sources with redshifts in the range $0.16<z<1.4$ and measurements of radio core dominance, a radio orientation indicator.  Measured properties include monochromatic luminosities at a range of wavelengths between the infrared and X-rays, integrated infrared luminosity, spectral slopes, and the covering fraction of the obscuring circumnuclear dust.  We estimate dust covering fraction assuming that the accretion disk emits anisotropically and discuss the shortcomings and technical difficulties of this calculation.  Luminosities are found to depend on orientation, with face-on sources factors of a $2-3$ brighter than more edge-on sources, depending on wavelength.  The degree of anisotropy varies very little with wavelength such that the overall shape of the spectral energy distribution does not vary significantly with orientation.  In the infrared, we do not observe a decrease in anisotropy with increasing wavelength.  The spectral slopes and estimates of covering fraction are not significantly orientation dependent.  We construct composite spectral energy distributions as a function of radio core dominance and find that these illustrate the results determined from the measured properties.
\end{abstract}


\section{introduction}
Quasars are notable for emitting their substantial luminosity over the entire electromagnetic spectrum.  A variety of physical processes, both thermal and non-thermal, and environments, with gas at a range of temperatures and distances from the central engine, are required to produce the observed quasar spectral energy distributions \citep[SEDs, for a review see][]{wilkes04}.  Quasars are too distant to be resolved spatially (except the radio jets that can be seen on the largest scales), so information about their internal structures must be gathered indirectly.  As a result, observed properties of the SED are used to infer much about the physical nature of quasars and a good understanding of the orientation dependence of these observations is necessary.

Generally speaking, all quasar SEDs exhibit similar emission components.  The most prominent feature is the peak in emission at optical/ultraviolet (UV) wavelengths, called the ``Big Blue Bump.''  This feature can be produced by thermal emission from the accretion disk surrounding the central supermassive black hole, where gas attains a range of temperatures, typically around $\sim10^{4-5}$~K.  

A corona surrounding the accretion disk is capable of up-scattering photons to X-ray energies, producing one component of the observed high-energy SED.  Radio-loud (RL) quasars have an additional jet-linked component to their X-ray emission \citep[e.g.,][]{kembhavi86}. 

The mid-infrared (MIR) portion of the SED is thought to be produced by a body of circumnuclear dust, hereafter referred to as a dusty torus.  The dust intercepts some fraction of the optical/UV emission from the accretion disk, reprocesses it, and re-emits at a range of wavelengths in the IR.  The torus is central to the unification of active galactic nuclei (AGN) that are observed with both broad and narrow emission lines (Type 1) and those that show only narrow emission lines (Type 2) \citep{antonucci93}.  In this picture, emission from the narrow line region is observed in both classes of AGN, but Type 1 sources are seen more face-on with a line of sight down into the central nucleus of the AGN where the broad lines are emitted.  Type 2 sources are seen more edge-on and the dusty torus blocks this view.

The radio region of the SED, which is primarily produced by a jet, is where RL and radio-quiet (RQ) sources distinguish themselves.  For similar optical emission, RL quasars can be many orders of magnitude brighter in the radio than RQ quasars.

The only well-established method of determining orientation in quasars depends on the anisotropy of the radio emission.  Radio orientation indicators rely on the beaming of the radio jet emission to estimate orientation \citep[e.g.,][]{willsbro95}.  Ideally, an anisotropically beamed quantity is normalized to one that is known to be emitted more isotropically, thus indicating the intrinsic power of the source.  Radio core dominance, the ratio of the of the radio core luminosity to the radio lobe luminosity at rest-frame 5~GHz, is one such orientation indicator.  \citet{ghisellini93} demonstrate that the radio core dominance parameter, log$\,R=\textrm{log}\,\left(\frac{L_{core}}{L_{ext.}}\right)$, tracks the viewing angle obtained from high-resolution radio maps that show superluminal motion in the radio jets of approximately 40 objects \citep[see][for an updated figure]{willsbro95}.  This method is incredibly powerful, but the technique is limited by the fact that it can be applied to only the 10\% of AGN that are RL because it depends exclusively on observations of the radio structures.

At least in RL quasars, the behavior of the radio emission as a function of orientation is well understood.  In some other wavelength regimes, the degree of anisotropy with which the SED is emitted is either known or can be anticipated based on physical models.  In the optical/UV early radiative transfer modeling of geometrically thin accretion disks suggests that the flux may vary by a factors of a few between edge-on and face-on sources \citep[e.g.,][]{laor89}.  More recent results from \citet{nb10}, who use the accretion disk models of \citet{hubeny00} that account for limb darkening and relativistic effects, show a similar degree of anisotropy in the optical/UV for typical quasar input parameters. 

The emission from the dusty torus can also be reproduced with some success via radiative transfer modeling.  In early models, most notably those of \citet{pier92}, the dust is distributed smoothly throughout the torus.  This induces a significant orientation dependence, over an order of magnitude difference in the flux from edge-on and face-on sources at 10~$\mu$m.  The anisotropy arises from the fact that the torus is optically thick and the hot dust near the central nucleus of the AGN is revealed at more face-on inclinations, where it contributes significantly to the MIR emission.  Later models adopt a clumpy distribution for the dust \citep{nenkova08a}.  In this case, the inner faces of the optically thick dust clouds are illuminated but can be observed from a wider range of inclinations extending to more edge-on views.  As a result, the models of \citet{nenkova08a} are capable of producing anisotropic emission in the MIR, but the orientation dependence can vary depending on the input parameters and is generally much less significant than for the smooth torus models.  The clumpy torus models are generally considered to be more realistic than those with a smooth dust distribution and have been shown to reproduce the observed near-infrared (NIR) and MIR SED well, even at high spatial resolution \citep[e.g.,][]{ramos_almeida11,alonso-herrero11}.

The SED atlas of \citet{shang11}, which includes the most detailed and up-to-date SEDs for 85 AGN and a subsample designed for studying orientation, provides an ideal opportunity to investigate the orientation dependence of quasar emission at all wavelengths.  Some properties have already been investigated for this sample in previous work.  \citet{runnoe13a} demonstrate that the optical/UV monochromatic luminosities depend on radio core dominance.  The X-ray-to-optical spectral index, which indicates the relative strength of the X-ray emission compared to the Big Blue Bump, does not depend on orientation (Runnoe et al. submitted).  The aim of this work is to compile these known dependencies and provide a new analysis of the orientation dependencies of the IR emission and the covering fraction of the dusty torus as well as construct composite SEDs to give a full picture of the orientation dependence of quasar SEDs.  The orientation dependence of the radio is well established, so we use this dependence to indicate orientation rather than include it in our analysis.

This investigation is structured as follows: in Section~\ref{sec:data} we discuss the sample and measurements.  The presentation of known orientation dependencies as well as new analysis of the IR and the creation of composite SEDs is detailed in Section~\ref{sec:analysis}.  The results are discussed in the context of previous work in Section~\ref{sec:discussion} and we summarize this work in Section~\ref{sec:conclusion}.  Throughout this work we use a cosmology with $H_0 = 70$ km s$^{-1}$ Mpc$^{-1}$, $\Omega_{\Lambda} = 0.7$, and $\Omega_{m} = 0.3$.

\section{Sample, Data, and Measurements}
\label{sec:data}
For this investigation we use the RL subsample of the \citet{shang11} SED atlas.  This sample is ideal for investigating orientation dependencies: objects are selected to have similar extended radio luminosity, which is thought to be emitted isotropically.  As a result of this selection, objects included in the sample are intrinsically similar but viewed at different angles.  This isolates the orientation effects that are of interest and minimizes differences caused by other parameters that might affect the SED, increasing our ability to discern statistically significant differences in the SEDs and their properties as a function of orientation.  The blazars originally included by \citeauthor{shang11} have been removed.  The total sample of 52 objects, with redshifts in the range $0.16<z<1.4$, is discussed in \citet{runnoe13a}.

Radio through X-ray SEDs are available for all of the objects in this sample, with varying levels of completeness.  \citet{shang11} describes the construction of the SEDs in detail, but we briefly describe the source of relevant data here.

X-ray data are available for 34 out of 52 objects (65\%).  Observations are from the {\it Chandra X-ray Observatory}, {\it XMM-Newton}, or {\it ROSAT} and are tabulated from the literature in \citet{shang11}.  In order to quantify the shape of the X-ray spectra for inclusion in the SEDs, \citet{shang11} use a power law or broken power law of the form $f_{\nu}=f_{0}E^{\alpha}$, where $f_{0}$ is the flux density at 1 keV, $E$ is the energy in keV, and $\alpha$ is the power-law spectral index.  They tabulate these parameters and use errors on the X-ray spectral index to determine uncertainty in this part of the SED.

Optical and UV spectrophotometry were obtained quasi-simultaneously for the entirety of this sample.  The sample was targeted with the {\it Hubble Space Telescope} (HST) as part of an early HST program and those observations were followed, usually within weeks, by optical observations from {\it Kitt Peak National Observatory} or {\it McDonald Observatory}.

IR coverage for the full SED sample comes from the 2 Micron All Sky Survey \citep[2MASS;][]{skrutskie06} and the {\it Spitzer Space Telescope}.  The RL subsample is rather poorly represented with only 9 objects having {\it Spitzer} coverage so we instead tabulate IR data from the {\it Wide-field Infrared Survey Explorer} \citep[WISE;][]{wright10}.  WISE covers four bands (W$1-4$) at 3.4, 4.6, 12, and 22 $\mu$m and is available for 42 (81\%) objects in the RL sample.  Using sources in the full SED sample that have both WISE and {\it Spitzer} coverage, \citet{runnoe12b} verify that there are likely no significant variability issues associated with including the WISE data as emission varies by less than $5-10\%$ in the $4-5$ years between {\it Spitzer} and WISE.  

Radio fluxes at multiple frequencies are available for this sample, tabulated from high-quality surveys and the literature by \citet{shang11}.  Not all sources are resolved in the original SED radio data, so \citet{runnoe13a} tabulate additional radio core and extended fluxes from the literature in order to calculate radio core dominance.

\citet{shang11} have applied two corrections during the construction of the SEDs. First, there is a correction for host emission that can contaminate emission from the  AGN at optical and NIR wavelengths.  Second is the correction for Galactic reddening to the emission at optical-to-far-ultraviolet wavelengths.

Optical, UV, and X-ray monochromatic luminosities have already been measured for this sample and are presented in \citet{runnoe12a}.  Here we add measurements of the monochromatic luminosity at 3, 8, and 12 $\mu$m, calculated from magnitudes available in the WISE All-Sky Data Release.  There is a known discrepancy that arises between red and blue calibrator stars for the WISE magnitudes.  Described in detail in \citet{wright10}, we correct this issue by adding $+17\%$ and $-9\%$ in flux to the W3 and W4 bands, respectively.  We then k-correct the luminosities to 3, 8, and 12 $\mu$m following the formalism of \citet{stocke92} and using the observed W2, W3, and W4 fluxes.  IR spectra of quasars are typically well approximated by a power law, so in order to obtain the slope for the k-correction, we fit a power law to all four WISE bands in log space and verify visually that this is reasonable.

In addition to these monochromatic luminosities, we also calculate an integrated luminosity in the MIR, $L$(MIR).  Similar to \citet{ma13}, we integrate a power law between the first and fourth WISE points over MIR wavelengths in order to determine this quantity.  This practice is intended to determine the emission from hot dust, but contamination from other sources can be an issue.  We select the wavelength range of rest-frame $3-12$~$\mu$m in order to minimize contributions from the accretion disk or cool dust heated by star formation in the host galaxy.  Using the Type 1 SED from \citet{silva04}, \citet{maiolino07} determine that the monochromatic luminosity at 6.7~$\mu$m must be scaled up by a factor of $\sim 2.7$ in order to estimate the integrated IR luminosity in the range $4-10$~$\mu$m.  We find a smaller correction of 1.8 on average, but for a different wavelength region. 

We measure spectral indices between the X-ray and optical and the optical and IR for this sample.  The slope between the flux at 2500~\AA\ and 2~keV, $\alpha_{ox}$, has already been measured for this sample by Runnoe et al. (submitted) and we calculate the slope between the flux at 2500~\AA\ and 12~$\mu$m, $\alpha_{oIR}$.  Both spectral indices are calculated directly from the data using the expression in \citet{steffen06} with appropriate fluxes and frequencies.   

We estimate covering fraction using a method similar to \citet{dipompeo13a}.  \citet{calderone12} derive the following limits on the covering fraction of the dusty torus (their equation~10):

\begin{eqnarray}
\label{eqn:CF}
\frac{c^2}{1+c} < R < c^2,
\end{eqnarray}

\noindent where $c$ is the covering fraction and $R$ is the observable $L(\textrm{MIR})/L_{bol}$.  This estimate accounts for the fact that the accretion disk luminosity is anisotropically emitted, although only via a geometric $\cos (\theta)$ effect.  We determine covering fraction estimates for our sample by inverting Equation~\ref{eqn:CF} to give limits on the covering fraction as a function of $R$ and averaging these upper and lower limits.  Bolometric luminosity is taken from \citet{runnoe12a} and is determined by integrating the SED between 1~$\mu$m and 2~keV or, when the complete SED is not observed, from the recommended bolometric correction at 1450 \AA\ derived in that work.  By excluding the IR emission in the bolometric luminosity integration, we avoid issues of double counting the IR photons.

Radio orientation indicators for this sample are presented in \citet{runnoe13a}.  Radio core dominance is calculated by k-correcting the luminosities from the radio core and extended radio lobes to rest-frame 5~GHz assuming slopes of $\alpha=0$ and $\alpha=-0.7$, respectively.  A measurement of radio core dominance requires that the source be resolved in the radio map.  The power of this method is that the core and lobe luminosities can be properly ascribed to their respective sources with reasonable certainty.

Table~\ref{tab:measurements} contains all luminosities, spectral indices, radio core dominance, and covering fraction measurements for this sample.


\onecolumn 
{\centering
\begin{landscape}
{\footnotesize
\renewcommand{\thefootnote}{\alph{footnote}}
\begin{ThreePartTable}
\begin{longtable}{lcccccccccrrrc}
\caption{Measurements \label{tab:measurements}}\\
Object
 & 
L$_{bol}$
 & 
$\,\lambda L_{\lambda}$(2~keV)
 & 
$\,\lambda L_{\lambda}$(1450~\AA)
 & 
$\,\lambda L_{\lambda}$(2500~\AA)
 & 
$\,\lambda L_{\lambda}$(5100~\AA)
 & 
$\,\lambda L_{\lambda}$(3~$\mu$m)
 & 
$\,\lambda L_{\lambda}$(8~$\mu$m)
 & 
$\,\lambda L_{\lambda}$(12~$\mu$m)
 & 
L(MIR)
 & 
$\alpha_{ox}$
 & 
$\alpha_{oIR}$
 & 
log$\,R$
 & 
$c$
 \\ 
      &ergs s$^{-1}$ & ergs s$^{-1}$ & ergs s$^{-1}$ & ergs s$^{-1}$ & ergs s$^{-1}$& ergs s$^{-1}$ & ergs s$^{-1}$ & ergs s$^{-1}$ & ergs s$^{-1}$ && & & \% \\
\hline
\endfirsthead
\multicolumn{14}{c}%
{\tablename\ \thetable\ -- \textit{Continued}} \\
Object
 & 
L$_{bol}$
 & 
$\,\lambda L_{\lambda}$(2~keV)
 & 
$\,\lambda L_{\lambda}$(1450~\AA)
 & 
$\,\lambda L_{\lambda}$(2500~\AA)
 & 
$\,\lambda L_{\lambda}$(5100~\AA)
 & 
$\,\lambda L_{\lambda}$(3~$\mu$m)
 & 
$\,\lambda L_{\lambda}$(8~$\mu$m)
 & 
$\,\lambda L_{\lambda}$(12~$\mu$m)
 & 
L(MIR)
 & 
$\alpha_{ox}$
 & 
$\alpha_{oIR}$
 & 
log$\,R$
 & 
$c$
 \\ 
      &ergs s$^{-1}$ & ergs s$^{-1}$ & ergs s$^{-1}$ & ergs s$^{-1}$ & ergs s$^{-1}$& ergs s$^{-1}$ & ergs s$^{-1}$ & ergs s$^{-1}$ & ergs s$^{-1}$ && & & \% \\
\hline
\endhead
\hline 
\multicolumn{14}{r}{\textit{Continued on next page}} \\
\endfoot
\hline
\endlastfoot

\hline
3C 110   
 & 
   46.99
 & 
\nodata
 & 
   46.43
 & 
   46.24
 & 
   45.99
 & 
\nodata
 & 
\nodata
 & 
\nodata
 & 
\nodata
 & 
\nodata
 & 
\nodata
 & 
$-$0.249
 & 
\nodata
 \\ 
3C 175   
 & 
   46.91
 & 
\nodata
 & 
   46.34
 & 
   46.31
 & 
   46.11
 & 
   45.88
 & 
   45.73
 & 
   45.74
 & 
   46.03
 & 
\nodata
 & 
$-$0.66
 & 
$-$1.836
 & 
          40
 \\ 
3C 186   
 & 
   46.66
 & 
\nodata
 & 
   46.07
 & 
   45.97
 & 
   45.79
 & 
   45.67
 & 
   45.71
 & 
   45.68
 & 
   45.94
 & 
\nodata
 & 
$-$0.83
 & 
   0.060
 & 
          48
 \\ 
3C 207   
 & 
   46.28
 & 
\nodata
 & 
   45.65
 & 
   45.65
 & 
   45.48
 & 
   45.31
 & 
   45.44
 & 
   45.43
 & 
   45.68
 & 
\nodata
 & 
$-$0.87
 & 
$-$0.393
 & 
          56
 \\ 
3C 215   
 & 
   45.77
 & 
   44.83
 & 
   45.02
 & 
   44.93
 & 
   44.94
 & 
   44.96
 & 
   44.88
 & 
   45.00
 & 
   45.26
 & 
$-$1.04
 & 
$-$1.04
 & 
$-$1.521
 & 
          64
 \\ 
3C 232   
 & 
   46.36
 & 
   44.50
 & 
   45.87
 & 
   45.94
 & 
   45.72
 & 
   45.36
 & 
   45.33
 & 
   45.38
 & 
   45.66
 & 
$-$1.55
 & 
$-$0.67
 & 
   0.641
 & 
          50
 \\ 
3C 254   
 & 
   46.17
 & 
   44.10
 & 
   45.63
 & 
   45.55
 & 
   45.36
 & 
\nodata
 & 
\nodata
 & 
\nodata
 & 
\nodata
 & 
$-$1.56
 & 
\nodata
 & 
$-$1.722
 & 
\nodata
 \\ 
3C 263   
 & 
   46.81
 & 
   43.75
 & 
   46.33
 & 
   46.23
 & 
   46.01
 & 
   45.92
 & 
   45.84
 & 
   45.86
 & 
   46.13
 & 
$-$1.95
 & 
$-$0.78
 & 
$-$0.912
 & 
          51
 \\ 
3C 277.1   
 & 
   45.61
 & 
   44.30
 & 
   45.04
 & 
   44.89
 & 
   44.67
 & 
   44.65
 & 
   44.86
 & 
   44.88
 & 
   45.11
 & 
$-$1.23
 & 
$-$0.99
 & 
$-$1.425
 & 
          65
 \\ 
3C 281   
 & 
   46.23
 & 
   45.06
 & 
   45.68
 & 
   45.58
 & 
   45.32
 & 
   45.31
 & 
   45.14
 & 
   45.18
 & 
   45.47
 & 
$-$1.20
 & 
$-$0.76
 & 
$-$0.967
 & 
          46
 \\ 
3C 288.1   
 & 
   46.61
 & 
\nodata
 & 
   46.01
 & 
   45.90
 & 
   45.62
 & 
   45.55
 & 
   45.63
 & 
   45.77
 & 
   46.01
 & 
\nodata
 & 
$-$0.92
 & 
$-$2.169
 & 
          56
 \\ 
3C 334   
 & 
   46.44
 & 
   44.55
 & 
   46.00
 & 
   45.90
 & 
   45.58
 & 
   45.58
 & 
   45.58
 & 
   45.74
 & 
   45.98
 & 
$-$1.52
 & 
$-$0.90
 & 
$-$0.648
 & 
          68
 \\ 
3C 37   
 & 
   46.18
 & 
   45.02
 & 
   45.28
 & 
   45.26
 & 
   44.89
 & 
   45.50
 & 
   45.44
 & 
   45.41
 & 
   45.69
 & 
$-$1.09
 & 
$-$1.09
 & 
$-$0.433
 & 
          66
 \\ 
3C 446   
 & 
   47.01
 & 
   46.25
 & 
   46.19
 & 
   46.28
 & 
   46.43
 & 
\nodata
 & 
\nodata
 & 
\nodata
 & 
\nodata
 & 
$-$1.01
 & 
\nodata
 & 
   1.528
 & 
\nodata
 \\ 
3C 47   
 & 
   45.97
 & 
   44.80
 & 
   45.27
 & 
   45.10
 & 
   44.88
 & 
   45.50
 & 
   45.47
 & 
   45.46
 & 
   45.73
 & 
$-$1.11
 & 
$-$1.22
 & 
$-$1.286
 & 
          92
 \\ 
4C 01.04   
 & 
   45.44
 & 
   44.56
 & 
   44.43
 & 
   44.56
 & 
   44.69
 & 
   44.75
 & 
   44.92
 & 
   44.93
 & 
   45.17
 & 
$-$1.00
 & 
$-$1.22
 & 
$-$0.621
 & 
          90
 \\ 
4C 06.69   
 & 
   47.30
 & 
   46.32
 & 
   46.73
 & 
   46.59
 & 
   46.43
 & 
   46.47
 & 
   46.48
 & 
   46.47
 & 
   46.72
 & 
$-$1.11
 & 
$-$0.93
 & 
   2.101
 & 
          58
 \\ 
4C 10.06   
 & 
   46.48
 & 
   45.30
 & 
   45.85
 & 
   45.73
 & 
   45.47
 & 
   45.40
 & 
   45.36
 & 
   45.33
 & 
   45.61
 & 
$-$1.17
 & 
$-$0.76
 & 
$-$0.658
 & 
          39
 \\ 
4C 12.40   
 & 
   46.17
 & 
\nodata
 & 
   45.53
 & 
   45.34
 & 
   45.13
 & 
   45.07
 & 
   45.03
 & 
   45.05
 & 
   45.33
 & 
\nodata
 & 
$-$0.83
 & 
$-$1.163
 & 
          41
 \\ 
4C 19.44   
 & 
   46.91
 & 
   45.76
 & 
   46.30
 & 
   46.21
 & 
   45.99
 & 
   45.90
 & 
   45.96
 & 
   46.05
 & 
   46.29
 & 
$-$1.17
 & 
$-$0.90
 & 
   0.427
 & 
          55
 \\ 
4C 20.24   
 & 
   46.92
 & 
   45.81
 & 
   46.22
 & 
   46.17
 & 
   45.95
 & 
   45.98
 & 
   46.01
 & 
   46.03
 & 
   46.28
 & 
$-$1.14
 & 
$-$0.92
 & 
$-$0.499
 & 
          54
 \\ 
4C 22.26   
 & 
   46.59
 & 
   44.91
 & 
   45.83
 & 
   45.70
 & 
   45.52
 & 
   45.44
 & 
   45.61
 & 
   45.62
 & 
   45.85
 & 
$-$1.30
 & 
$-$0.95
 & 
$-$1.028
 & 
          47
 \\ 
4C 30.25   
 & 
   46.27
 & 
\nodata
 & 
   45.64
 & 
   45.53
 & 
   45.13
 & 
   45.28
 & 
   45.31
 & 
   45.27
 & 
   45.54
 & 
\nodata
 & 
$-$0.84
 & 
$-$1.740
 & 
          48
 \\ 
4C 31.63   
 & 
   46.61
 & 
   45.00
 & 
   46.21
 & 
   46.03
 & 
   45.61
 & 
   45.58
 & 
   45.54
 & 
   45.46
 & 
   45.75
 & 
$-$1.39
 & 
$-$0.66
 & 
   0.979
 & 
          40
 \\ 
4C 39.25   
 & 
   46.91
 & 
   45.63
 & 
   46.25
 & 
   46.15
 & 
   45.97
 & 
   45.88
 & 
   45.88
 & 
   45.83
 & 
   46.09
 & 
$-$1.20
 & 
$-$0.81
 & 
$-$2.220
 & 
          43
 \\ 
4C 40.24   
 & 
   46.60
 & 
   45.47
 & 
   45.96
 & 
   45.87
 & 
   45.74
 & 
   45.83
 & 
   45.96
 & 
   45.78
 & 
   46.05
 & 
$-$1.15
 & 
$-$0.95
 & 
$-$0.477
 & 
          60
 \\ 
4C 41.21   
 & 
   46.75
 & 
   45.42
 & 
   46.27
 & 
   46.01
 & 
   45.64
 & 
   45.84
 & 
   45.75
 & 
   45.70
 & 
   45.97
 & 
$-$1.23
 & 
$-$0.81
 & 
$-$0.477
 & 
          45
 \\ 
4C 49.22   
 & 
   45.99
 & 
   44.96
 & 
   45.21
 & 
   45.14
 & 
   44.84
 & 
   45.47
 & 
   45.51
 & 
   45.49
 & 
   45.75
 & 
$-$1.07
 & 
$-$1.21
 & 
$-$0.046
 & 
          92
 \\ 
4C 55.17   
 & 
   46.46
 & 
   45.16
 & 
   45.83
 & 
   45.78
 & 
   45.72
 & 
   45.92
 & 
   46.00
 & 
   45.98
 & 
   46.24
 & 
$-$1.24
 & 
$-$1.12
 & 
   1.238
 & 
          95
 \\ 
4C 58.29   
 & 
   47.20
 & 
   45.09
 & 
   46.73
 & 
   46.58
 & 
   46.42
 & 
   46.18
 & 
   46.21
 & 
   46.16
 & 
   46.42
 & 
$-$1.57
 & 
$-$0.76
 & 
$-$1.561
 & 
          45
 \\ 
4C 64.15   
 & 
   46.73
 & 
\nodata
 & 
   46.14
 & 
   46.03
 & 
   45.98
 & 
\nodata
 & 
\nodata
 & 
\nodata
 & 
\nodata
 & 
\nodata
 & 
\nodata
 & 
$-$1.793
 & 
\nodata
 \\ 
4C 73.18   
 & 
   46.40
 & 
   44.98
 & 
   45.80
 & 
   45.77
 & 
   45.55
 & 
   45.63
 & 
   45.57
 & 
   45.59
 & 
   45.87
 & 
$-$1.30
 & 
$-$0.89
 & 
   1.265
 & 
          62
 \\ 
B2 0742+31   
 & 
   46.46
 & 
   45.34
 & 
   45.89
 & 
   45.93
 & 
   45.77
 & 
   45.86
 & 
   45.75
 & 
   45.83
 & 
   46.10
 & 
$-$1.22
 & 
$-$0.94
 & 
   0.360
 & 
          78
 \\ 
B2 1351+31   
 & 
   46.66
 & 
\nodata
 & 
   46.07
 & 
   45.94
 & 
   45.81
 & 
   45.53
 & 
   45.65
 & 
   45.63
 & 
   45.87
 & 
\nodata
 & 
$-$0.81
 & 
$-$0.474
 & 
          44
 \\ 
B2 1555+33   
 & 
   46.21
 & 
\nodata
 & 
   45.57
 & 
   45.44
 & 
   45.39
 & 
   45.44
 & 
   45.39
 & 
   45.27
 & 
   45.57
 & 
\nodata
 & 
$-$0.90
 & 
   0.212
 & 
          54
 \\ 
B2 1611+34   
 & 
   47.06
 & 
\nodata
 & 
   46.50
 & 
   46.35
 & 
   46.29
 & 
   46.03
 & 
   46.11
 & 
   46.16
 & 
   46.41
 & 
\nodata
 & 
$-$0.89
 & 
   0.014
 & 
          53
 \\ 
MC2 0042+101   
 & 
   45.68
 & 
\nodata
 & 
   44.99
 & 
   45.04
 & 
   44.94
 & 
   44.94
 & 
   44.81
 & 
   44.72
 & 
   45.04
 & 
\nodata
 & 
$-$0.81
 & 
$-$0.821
 & 
          54
 \\ 
MC2 1146+111   
 & 
   46.29
 & 
\nodata
 & 
   45.66
 & 
   45.71
 & 
   45.52
 & 
   45.41
 & 
   45.29
 & 
   45.05
 & 
   45.39
 & 
\nodata
 & 
$-$0.60
 & 
$-$1.311
 & 
          38
 \\ 
OS 562   
 & 
   46.74
 & 
   45.03
 & 
   46.21
 & 
   46.03
 & 
   45.82
 & 
   45.64
 & 
   45.64
 & 
   45.65
 & 
   45.92
 & 
$-$1.38
 & 
$-$0.78
 & 
   0.891
 & 
          43
 \\ 
PG 1100+772   
 & 
   46.46
 & 
   45.25
 & 
   45.90
 & 
   45.74
 & 
   45.44
 & 
   45.43
 & 
   45.34
 & 
   45.32
 & 
   45.61
 & 
$-$1.19
 & 
$-$0.75
 & 
$-$1.095
 & 
          41
 \\ 
PG 1103-006   
 & 
   46.30
 & 
\nodata
 & 
   45.67
 & 
   45.61
 & 
   45.30
 & 
\nodata
 & 
\nodata
 & 
\nodata
 & 
\nodata
 & 
\nodata
 & 
\nodata
 & 
$-$0.391
 & 
\nodata
 \\ 
PG 1226+023   
 & 
   46.96
 & 
   45.70
 & 
   46.45
 & 
   46.34
 & 
   45.97
 & 
   45.96
 & 
   45.78
 & 
   45.74
 & 
   46.05
 & 
$-$1.25
 & 
$-$0.64
 & 
   0.643
 & 
          38
 \\ 
PG 1545+210   
 & 
   45.93
 & 
   44.62
 & 
   45.39
 & 
   45.32
 & 
   45.07
 & 
   45.22
 & 
   45.05
 & 
   45.05
 & 
   45.35
 & 
$-$1.27
 & 
$-$0.84
 & 
$-$1.750
 & 
          58
 \\ 
PG 1704+608   
 & 
   46.49
 & 
   44.74
 & 
   45.91
 & 
   45.86
 & 
   45.72
 & 
   45.85
 & 
   45.72
 & 
   45.81
 & 
   46.08
 & 
$-$1.43
 & 
$-$0.97
 & 
$-$1.385
 & 
          73
 \\ 
PG 2251+113   
 & 
   46.35
 & 
\nodata
 & 
   45.73
 & 
   45.73
 & 
   45.53
 & 
   45.59
 & 
   45.41
 & 
   45.37
 & 
   45.69
 & 
\nodata
 & 
$-$0.79
 & 
$-$2.501
 & 
          52
 \\ 
PKS 0112-017   
 & 
   46.88
 & 
   45.47
 & 
   46.44
 & 
   46.30
 & 
   46.07
 & 
\nodata
 & 
\nodata
 & 
\nodata
 & 
\nodata
 & 
$-$1.32
 & 
\nodata
 & 
$-$0.569
 & 
\nodata
 \\ 
PKS 0403-13   
 & 
   46.36
 & 
   45.51
 & 
   45.69
 & 
   45.65
 & 
   45.53
 & 
\nodata
 & 
\nodata
 & 
\nodata
 & 
\nodata
 & 
$-$1.05
 & 
\nodata
 & 
   1.696
 & 
\nodata
 \\ 
PKS 0859-14   
 & 
   47.22
 & 
\nodata
 & 
   46.68
 & 
   46.56
 & 
   46.45
 & 
\nodata
 & 
\nodata
 & 
\nodata
 & 
\nodata
 & 
\nodata
 & 
\nodata
 & 
   1.404
 & 
\nodata
 \\ 
PKS 1127-14   
 & 
   47.15
 & 
   46.31
 & 
   46.50
 & 
   46.40
 & 
   46.21
 & 
\nodata
 & 
\nodata
 & 
\nodata
 & 
\nodata
 & 
$-$1.04
 & 
\nodata
 & 
   1.950
 & 
\nodata
 \\ 
PKS 1656+053   
 & 
   47.06
 & 
   46.14
 & 
   46.44
 & 
   46.36
 & 
   46.26
 & 
   46.23
 & 
   46.29
 & 
   46.27
 & 
   46.51
 & 
$-$1.08
 & 
$-$0.94
 & 
$-$0.017
 & 
          61
 \\ 
PKS 2216-03   
 & 
   46.95
 & 
\nodata
 & 
   46.38
 & 
   46.39
 & 
   46.31
 & 
\nodata
 & 
\nodata
 & 
\nodata
 & 
\nodata
 & 
\nodata
 & 
\nodata
 & 
   0.922
 & 
\nodata
 \\ 
TEX 1156+213   
 & 
   46.01
 & 
\nodata
 & 
   45.36
 & 
   45.26
 & 
   45.00
 & 
   45.10
 & 
   45.00
 & 
   44.89
 & 
   45.20
 & 
\nodata
 & 
$-$0.78
 & 
$-$1.228
 & 
          43
 \\ 
\hline
\end{longtable}
\begin{tablenotes}
\small
 \item Note $-$ All luminosities are in log space.
\end{tablenotes}
\end{ThreePartTable}
} 
\end{landscape}
} 

\twocolumn

\section{Analysis}
\label{sec:analysis}
We investigate the orientation dependence of a variety of properties of the SED.  The analysis consists of three parts.  The first part is a rank correlation analysis of the orientation dependence of properties of the SED.  Second, we investigate in particular the orientation dependence of the covering fraction of the dusty torus.  Finally, we construct composite SEDs in order to demonstrate visually the differences in the SED with orientation.

\subsection{Properties of the SED}
\begin{figure*}
\begin{minipage}[!b]{7cm}
\centering
\includegraphics[width=7cm]{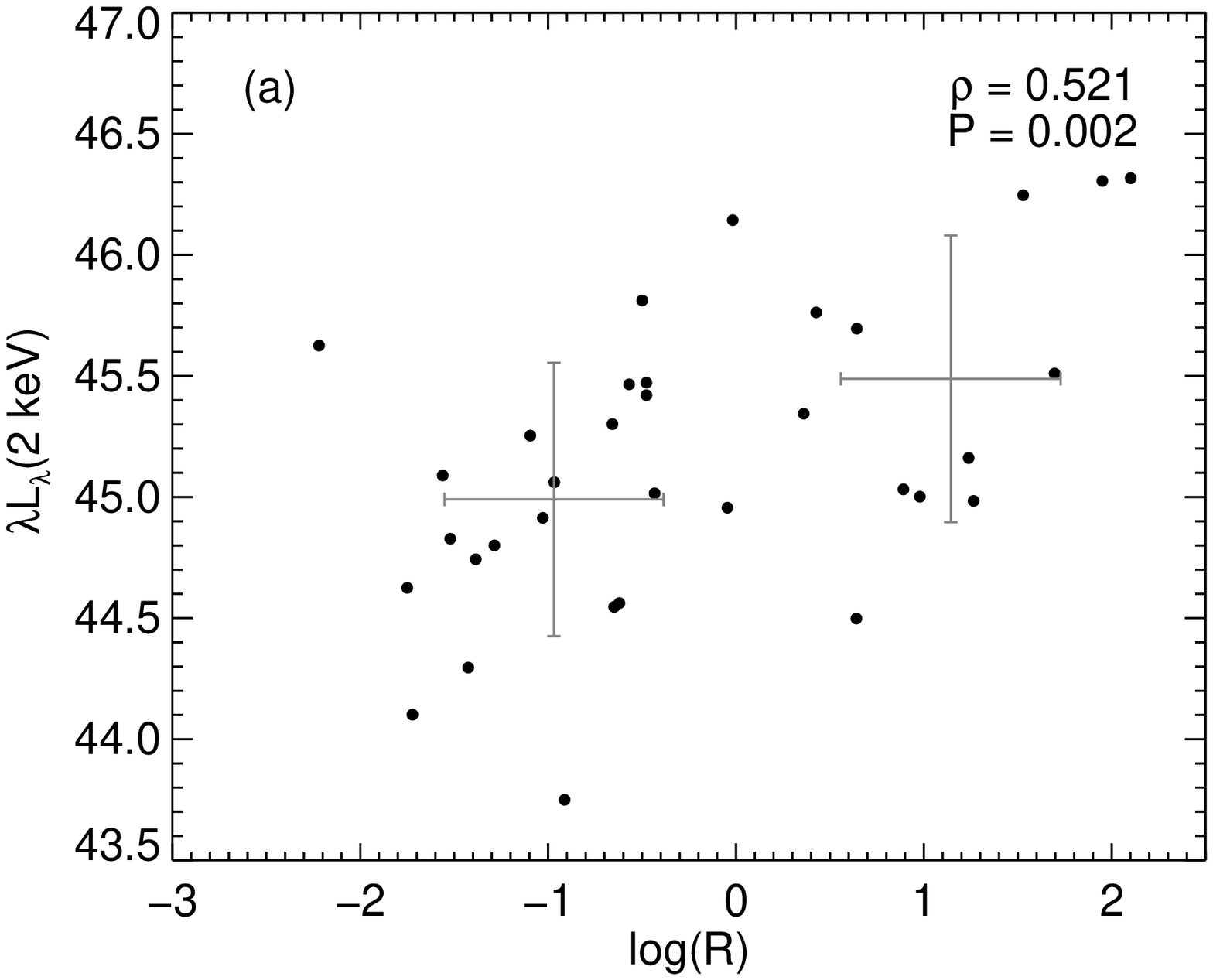}
\end{minipage}
\hspace{0.7cm}
\begin{minipage}[!b]{7cm}
\centering
\includegraphics[width=7cm]{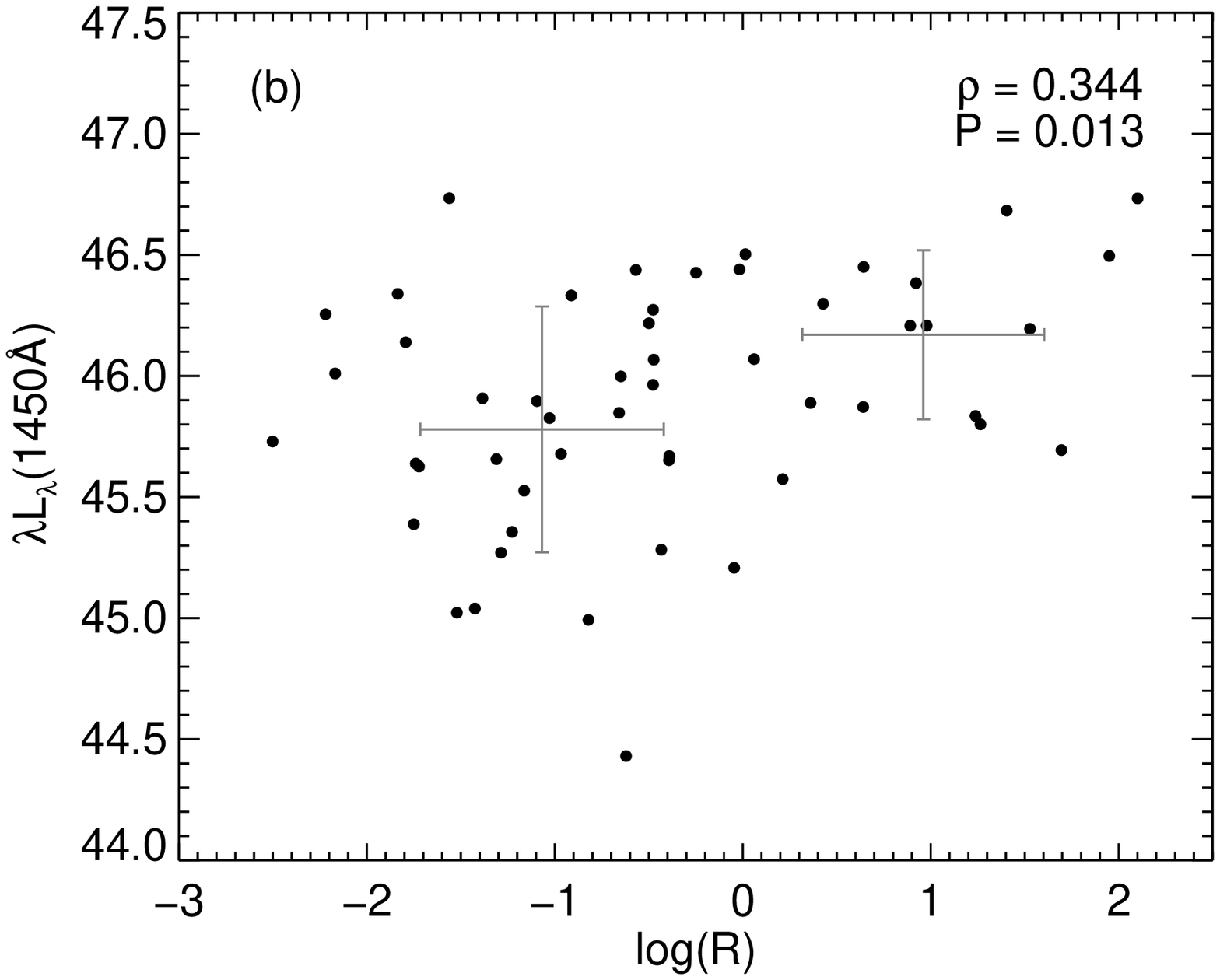}
\end{minipage}
\hspace{0.7cm}
\begin{minipage}[!b]{7cm}
\centering
\includegraphics[width=7cm]{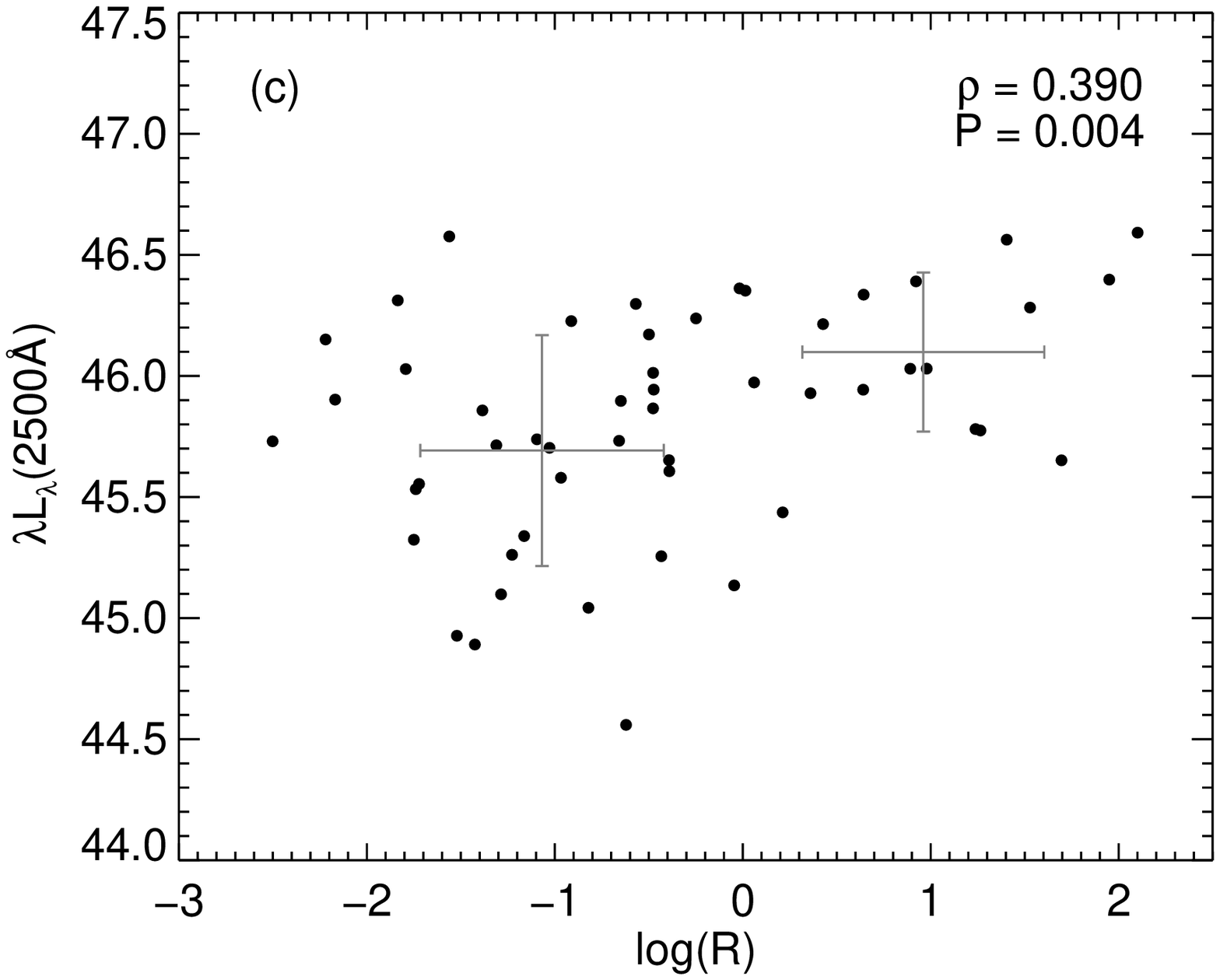}
\end{minipage}     
\hspace{0.7cm}       
\begin{minipage}[!b]{7cm}
\centering
\includegraphics[width=7cm]{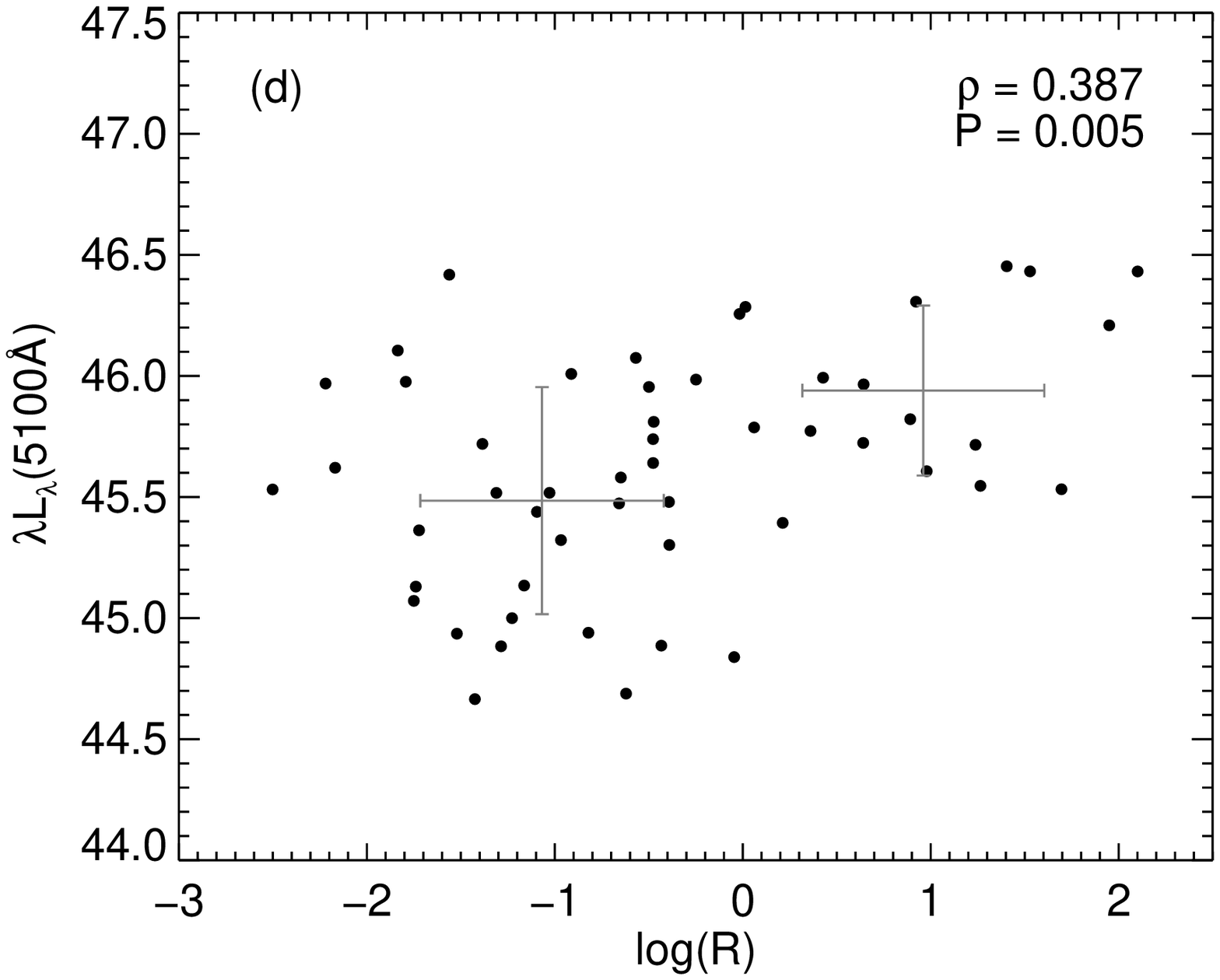}
\end{minipage}           
\begin{minipage}[!b]{7cm}
\centering
\includegraphics[width=7cm]{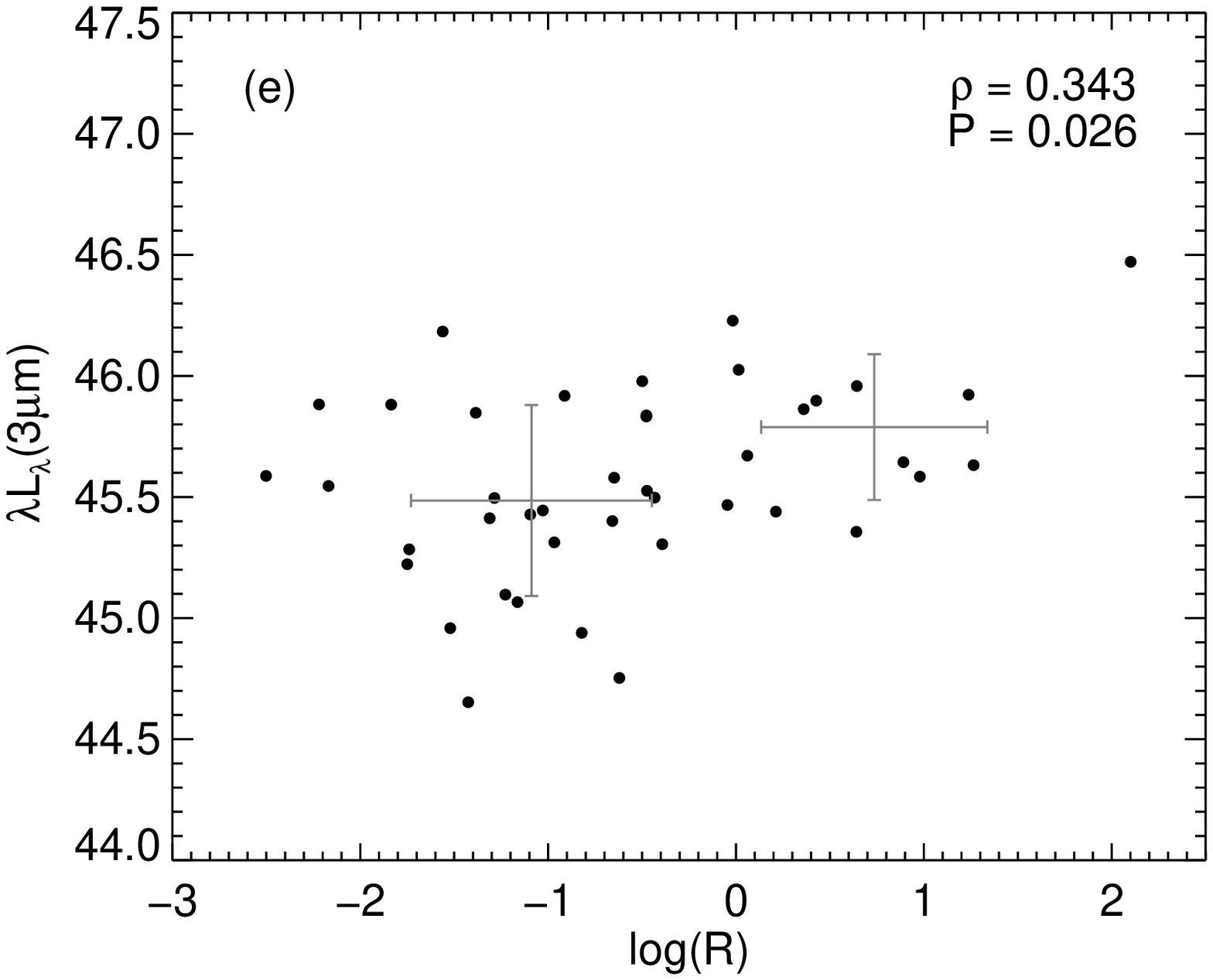}
\end{minipage}
\hspace{0.7cm}
\begin{minipage}[!b]{7cm}
\centering
\includegraphics[width=7cm]{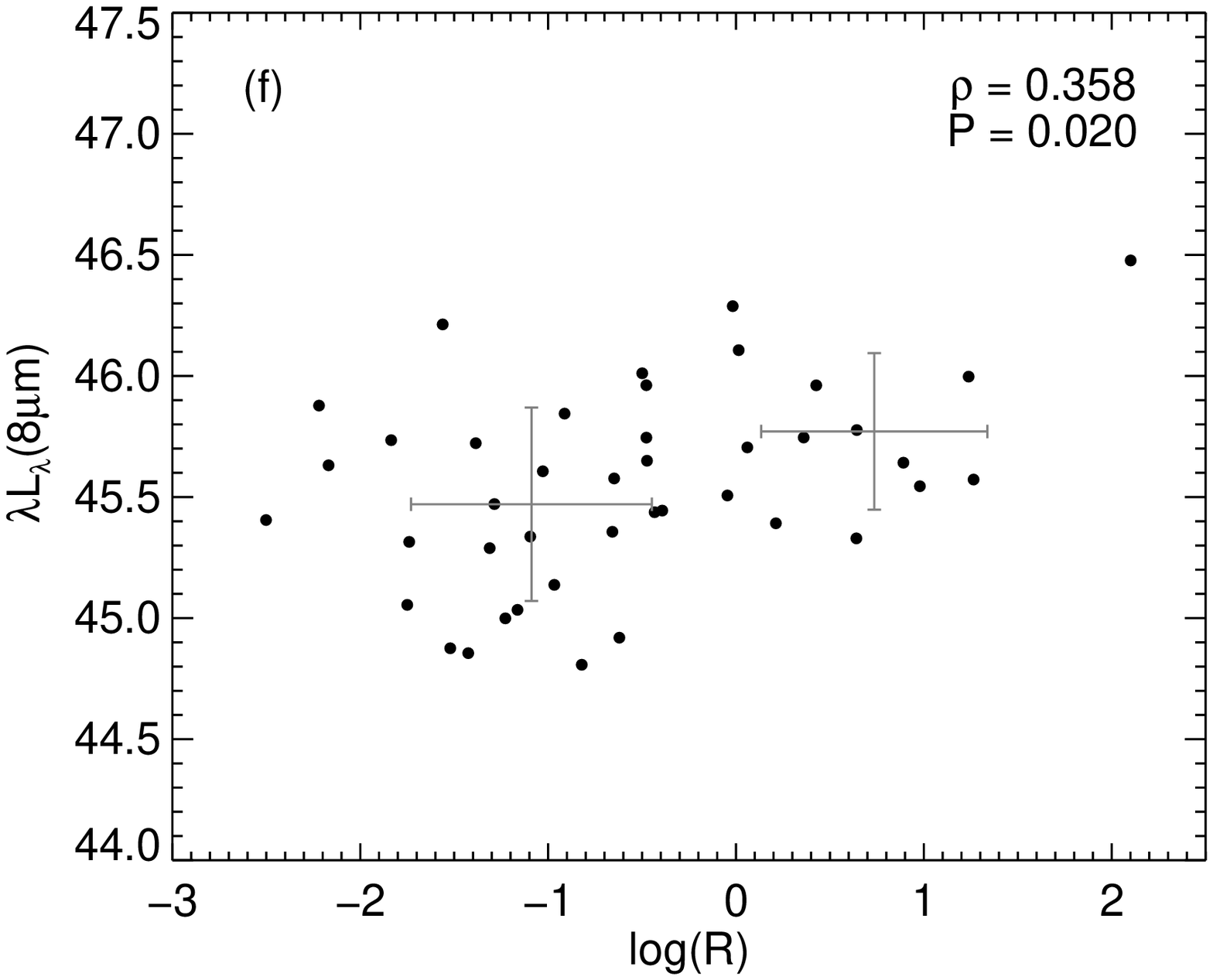}
\end{minipage}
\hspace{0.7cm}
\begin{minipage}[!b]{7cm}
\centering
\includegraphics[width=7cm]{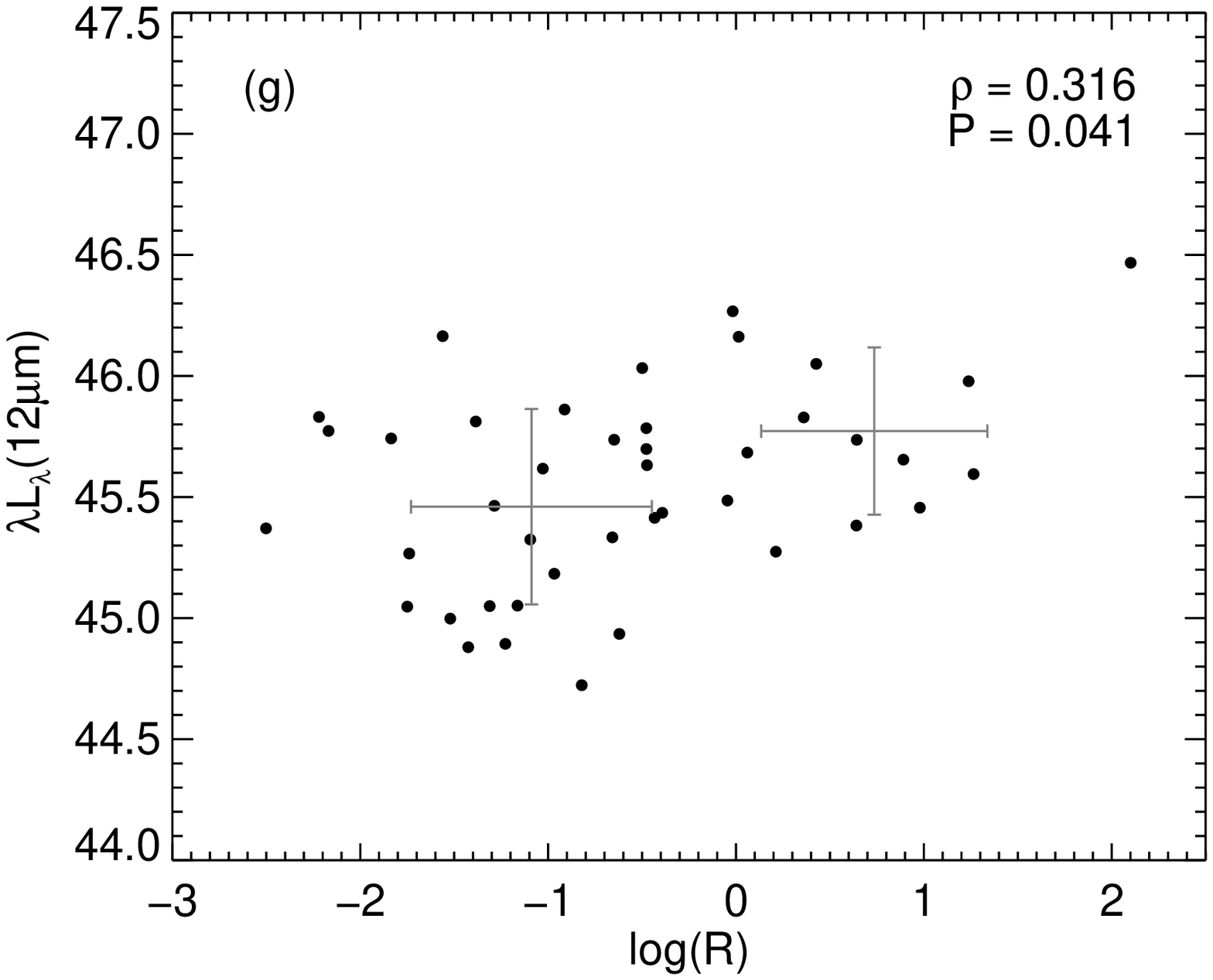}
\end{minipage}     
\hspace{0.7cm}       
\begin{minipage}[!b]{7cm}
\centering
\includegraphics[width=7cm]{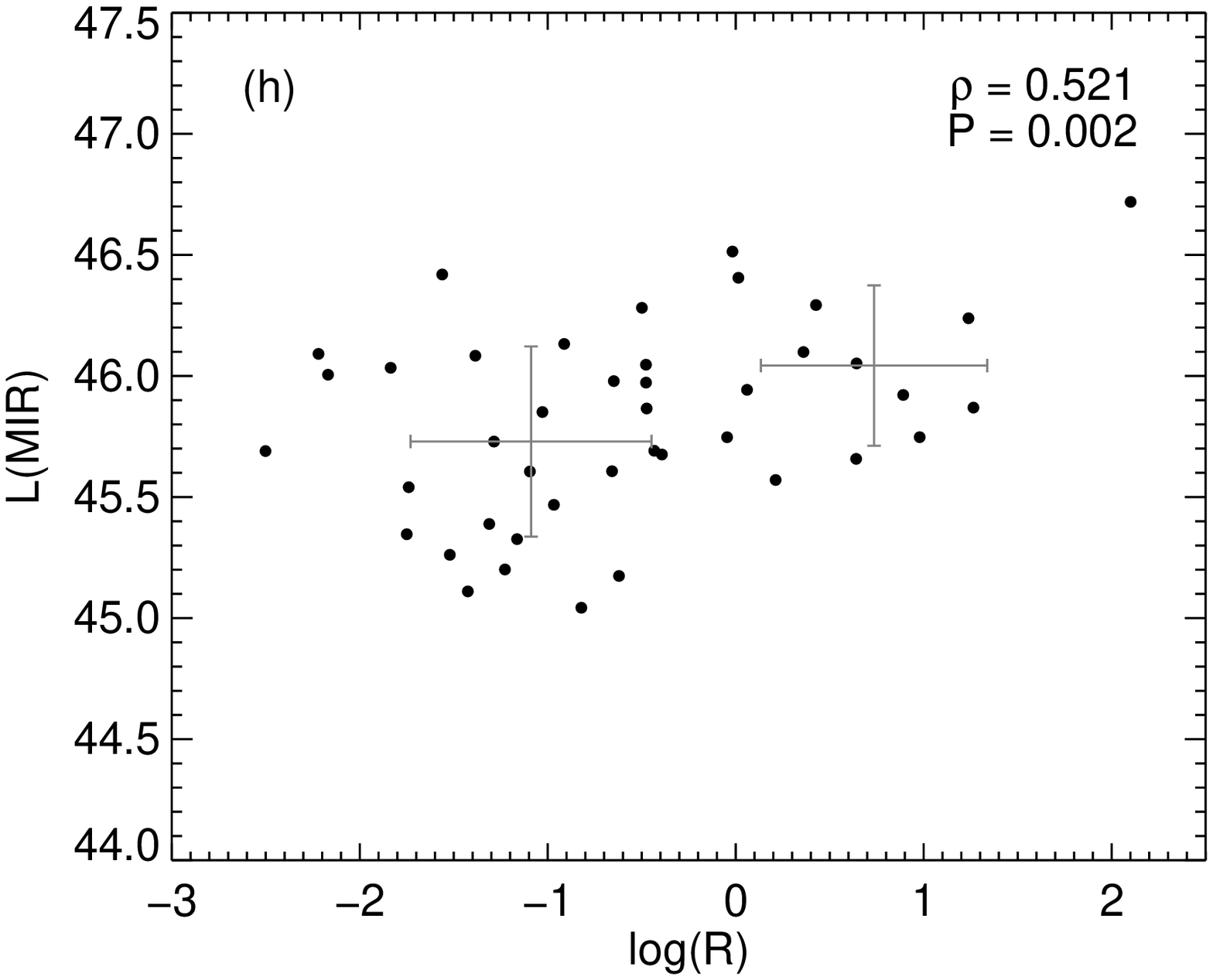}
\end{minipage}             
\caption{Monochromatic and integrated IR luminosities versus radio core dominance, an orientation indicator.  The Spearman Rank correlation coefficient and associated probability of finding these distributions of points by chance are listed in the upper right corner of each plot.  Gray crosses indicate the average values for core-dominated (log~$R\ge0$) and and lobe-dominated (log~$R<0$) sources with 1$\sigma$ dispersion in the X and Y directions.  The orientation dependence is marginally significant at 1450 \AA\ and at all IR wavelengths and significant at 2~keV, 2500 \AA, and 5100 \AA.  \label{fig:Lorient}}
\label{fig:corr}
\end{figure*}

\begin{figure*}
\begin{minipage}[!b]{8.cm}
\centering
\includegraphics[width=8.9cm]{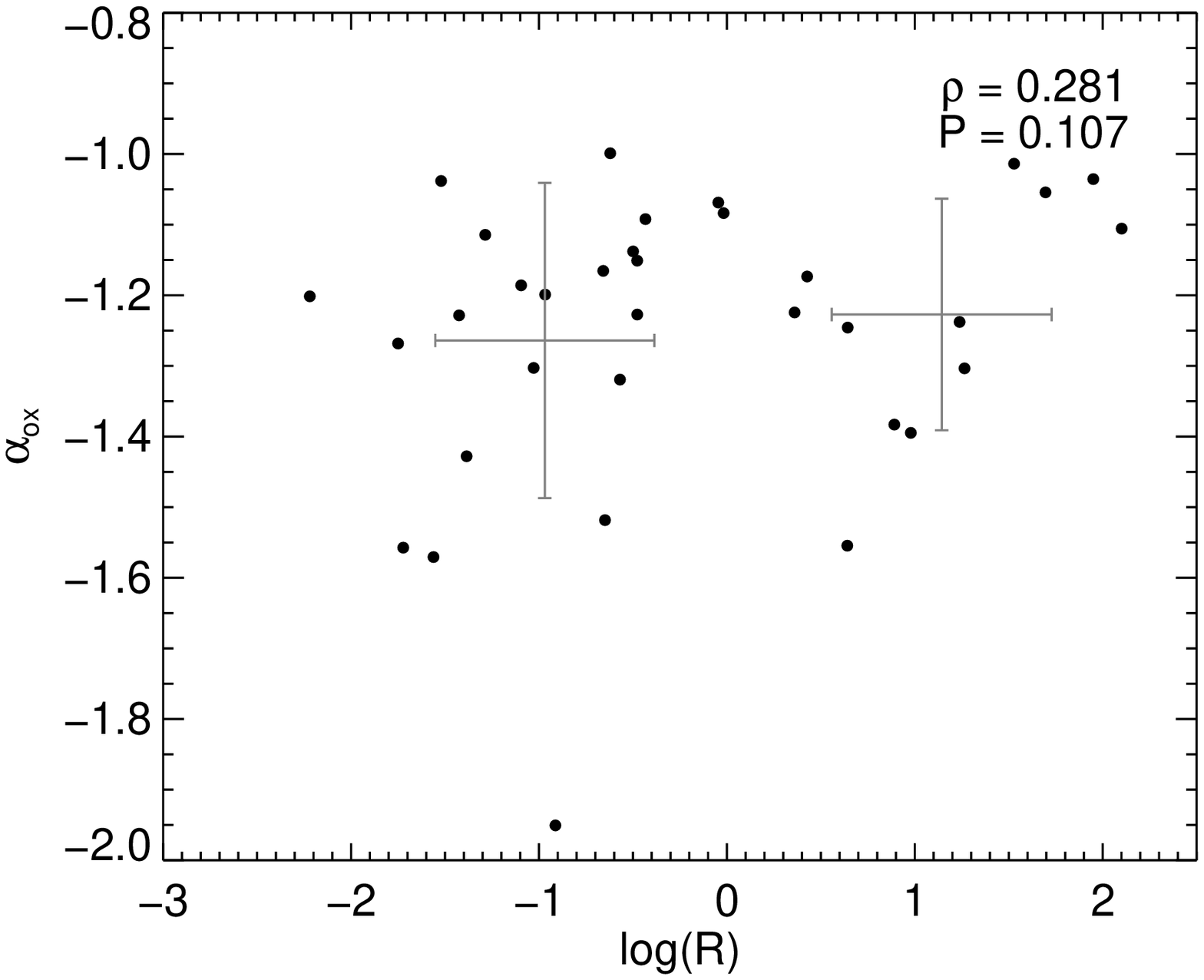}
\end{minipage}
\hspace{0.4cm}
\begin{minipage}[!b]{8.cm}
\centering
\includegraphics[width=8.9cm]{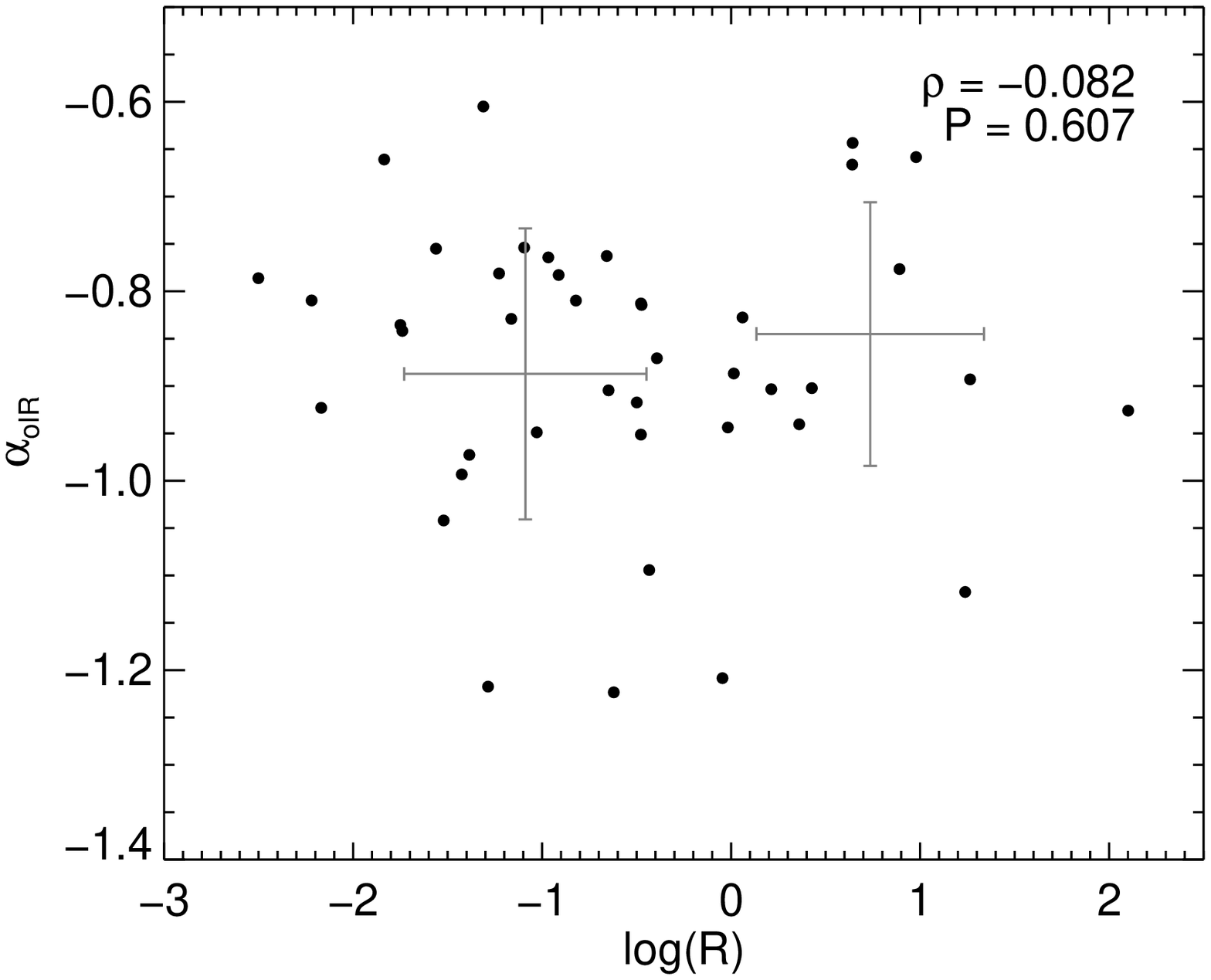}
\end{minipage}
\caption{The spectral slopes $\alpha_{ox}$ and $\alpha_{oIR}$ and radio core dominance.  Neither shows a statistically significant orientation dependence.  The Spearman Rank correlation coefficient and associated probability of finding these distributions of points by chance are listed in the upper right corner of each plot.  Gray crosses indicate the average values for core-dominated (log~$R\ge0$) and and lobe-dominated (log~$R<0$) sources with 1$\sigma$ dispersion in the X and Y directions.}
\label{fig:alphorient}
\end{figure*}

In order to quantify the orientation dependence of the measured luminosities and spectral slopes, we present the Spearman Rank (RS) correlations between these and radio core dominance in Table~\ref{tab:corr}.  For each correlation we tabulate the number of objects (which may be fewer than the total sample size if data are not available), the RS statistic, $\rho$, and the probability, $P$, of achieving the observed distribution of points by chance.  We define a significant correlation as one having a value of $P$ that is 0.01 or less and a marginally significant correlation as having a value of $P$ that is 0.05 or less.  In Table~\ref{tab:corr} these correlations, which we take to indicate an orientation dependence, are highlighted in bold.  The correlations between the luminosities and radio core dominance are illustrated in Figure~\ref{fig:Lorient} and those between spectral slopes and radio core dominance are illustrated in Figure~\ref{fig:alphorient}.  In these figures, average values of each parameter are presented with the 1$\sigma$ dispersion in the X and Y directions to indicate typical values for core-dominated (log~$R\ge0$) and and lobe-dominated (log~$R<0$) sources.  These average values are tabulated in Table~\ref{tab:avg} and indicate the average degree of anisotropy at each wavelength.

\begin{table}
\begin{minipage}[2cm]{6.75cm}
\renewcommand{\thefootnote}{\alph{footnote}}
\caption{Correlations with orientation \label{tab:corr}}
\begin{tabular}{llcrr}
Param 1 & Param 2 & N & $\rho$   & P \\
\hline
 $\lambda$L$_{\lambda}$(2 keV)  	& log~$R$    &           34& $   0.521$ & $\boldsymbol{   0.002}$ \\
 $\lambda$L$_{\lambda}$(1450 \AA)  	& log~$R$    &           52& $   0.344$ & $\boldsymbol{   0.013}$ \\
 $\lambda$L$_{\lambda}$(2500 \AA)  	& log~$R$    &           52& $   0.390$ & $\boldsymbol{   0.004}$ \\
 $\lambda$L$_{\lambda}$(5100 \AA)  	& log~$R$    &           52& $   0.387$ & $\boldsymbol{   0.005}$ \\
 $\lambda$L$_{\lambda}$(3 $\mu$m)  	& log~$R$    &           42 & $   0.343$ & $\boldsymbol{   0.026}$ \\
 $\lambda$L$_{\lambda}$(8 $\mu$m)  	& log~$R$    &           42 & $   0.358$ & $\boldsymbol{   0.020}$ \\
 $\lambda$L$_{\lambda}$(12 $\mu$m)  	& log~$R$    &           42 & $   0.316$ & $\boldsymbol{   0.041}$ \\
 L(MIR)  			  	& log~$R$    &           42 & $   0.336$ & $\boldsymbol{   0.030}$ \\
 $\alpha_{ox}$ 			& log~$R$    &           34 & $   0.281$ & $   0.107$ \\
 $\alpha_{oIR}$			& log~$R$    &           42 & $  -0.082$ & $   0.607$ \\
 $c$					& log~$R$    &           42 & $   0.141$ & $   0.373$ \\
\hline
\end{tabular}
\footnotetext[0]{Note $-$ Not all measurements are available in all sources, thus the number used in a correlation may be fewer than in the total sample.  The Spearman Rank correlation coefficient, $\rho$, and the probability that the observed distribution of points might arise by chance, $P$, are given in the table.  Significant and marginally significant correlations that indicate an orientation dependence are given in bold.}
\end{minipage}
\end{table}

\begin{table}
\begin{minipage}[2cm]{8.3cm}
\renewcommand{\thefootnote}{\alph{footnote}}
\caption{Average parameter values \label{tab:avg}}
\begin{tabular}{lrrc}
Parameter & Core & Lobe & Core/Lobe \\
\hline
 log$\,\lambda$L$_{\lambda}$(2 keV) &    45.49$\pm$0.59 &    44.99$\pm$0.56 &     3.15\\
 log$\,\lambda$L$_{\lambda}$(1450 \AA) &    46.17$\pm$0.35 &    45.78$\pm$0.51 &     2.46\\
 log$\,\lambda$L$_{\lambda}$(2500 \AA) &    46.10$\pm$0.33 &    45.69$\pm$0.48 &     2.55\\
 log$\,\lambda$L$_{\lambda}$(5100 \AA) &    45.94$\pm$0.35 &    45.49$\pm$0.47 &     2.85\\
 log$\,\lambda$L$_{\lambda}$(3 $\mu$m) &    45.79$\pm$0.30 &    45.49$\pm$0.39 &     2.01\\
 log$\,\lambda$L$_{\lambda}$(8 $\mu$m) &    45.77$\pm$0.32 &    45.47$\pm$0.40 &     2.00\\
 log$\,\lambda$L$_{\lambda}$(12 $\mu$m) &    45.77$\pm$0.35 &    45.46$\pm$0.40 &     2.05\\
 L(MIR) &     46.04$\pm$0.33 &    45.73$\pm$0.39 &     2.06\\
 $\alpha_{ox}$ & $-$1.23$\pm$0.16 & $-$1.26$\pm$0.22 &     0.97\\
 $\alpha_{oIR}$ & $-$0.85$\pm$0.14 & $-$0.89$\pm$0.15 &     0.95\\
 $c$ &      0.57$\pm$0.16 &     0.56$\pm$0.15 &     1.01\\
\hline
\end{tabular}
\footnotetext[0]{Note $-$ Average values of each parameter are for core-dominated (log~$R\ge0$) and lobe-dominated (log~$R<0$) sources.  Uncertainties are the 1$\sigma$ dispersion for sources in each bin.  The Core/Lobe ratio is in linear space.}
\end{minipage}
\end{table}

Of the monochromatic luminosities,the X-ray luminosity at 2~keV and the 2500 and 5100 \AA\ luminosities show a significant dependence on radio core dominance.  The UV and IR luminosities show marginally significant dependencies.  In general, face-on sources are brighter than more edge-on sources by factors of $2-3$ depending on wavelength, but the degree of anisotropy at each wavelength is such that the shape of the SED changes very little with orientation.  We also find that neither the optical-X-ray nor optical-IR spectral slopes show a statistically significant orientation dependence.  Some of these results are previously known and will be discussed further in Section~\ref{sec:discussion}.  

In the IR it is common to compare the orientation dependencies as a function of wavelength.  The significance of the correlation is similar at 3, 8, and 12~$\mu$m and, comparing the degree of anisotropy between the core-dominated and lobe-dominated sources, there is no significant change with wavelength.

\subsection{Covering fraction}
We have estimated the covering fraction from the ratio of the integrated luminosity in the MIR to the bolometric luminosity, what is essentially the fraction of the bolometric luminosity emitted by the accretion disk that is reprocessed into the IR by a dusty torus.  We find a mean covering fraction for the total sample of $0.56\pm0.15$ (corresponding to an opening angle from the jet axis of $56^\circ$), where the uncertainty is the standard deviation of the covering fractions measured in this sample.

\begin{figure}
\begin{center}
\includegraphics[width=8.9 truecm]{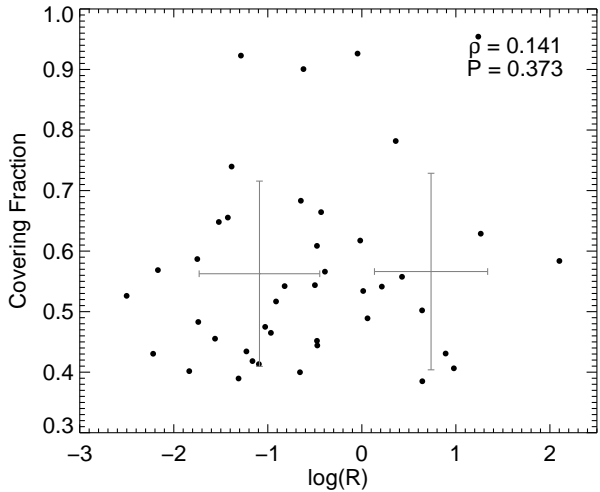}
\end{center}
\caption{The covering fraction of the dusty torus and radio core dominance.  The covering fraction does not significantly depend on orientation.  The Spearman Rank correlation coefficient and associated probability of finding these distributions of points by chance are listed in the upper right corner of the plot.  Gray crosses indicate the average values for core-dominated (log~$R\ge0$) and and lobe-dominated (log~$R<0$) sources with 1$\sigma$ dispersion in the X and Y directions.}
\label{fig:cforient}
\end{figure}

The correlation between radio core dominance and estimates of covering fraction is presented in Table~\ref{tab:corr} and illustrated in Figure~\ref{fig:cforient}.  We find that estimates of the covering fraction of the dusty torus do not correlate with radio core dominance.  This is not surprising given the similar orientation dependencies of the 12~$\mu$m and 1450~\AA\ luminosities, where the 1450 \AA\ luminosity is a proxy for the bolometric luminosity.  Though not identical, the orientation dependencies of the IR and UV luminosities are very similar.

\subsection{Composite SEDs}
Making composites is a powerful way of visually demonstrating the orientation dependence of the SED.  We create two composites for sources with log~$R<0$ and log~$R>0$ following the method of \citet{shang11}.  

The construction process is as follows.  Unlike \citet{shang11}, we do not normalize the SEDs before constructing the composites because the sample is selected on extended radio luminosity to include intrinsically similar objects with varying orientation.  The first step is to bin the data.  Each bin will contribute one point to the final SED at the center frequency of the bin and we use same bins for the creation of each composite SED.  Our bins are similar and sometimes identical to those used in \citet{shang11}.  We use the distribution of points for all objects to be included in the composite to manually determine the binning.  In the optical/UV and X-ray where the SEDs include continuous spectra, it is straightforward to create bins of uniform width across the relevant frequency range.  In the radio and IR where the data coverage is less consistent, bins are determined manually.  In these cases, we try to exclude regions without data and create similar width bins over the remaining frequency space.

\begin{figure*}
\begin{minipage}[!b]{8cm}
\centering
\includegraphics[width=8.9cm]{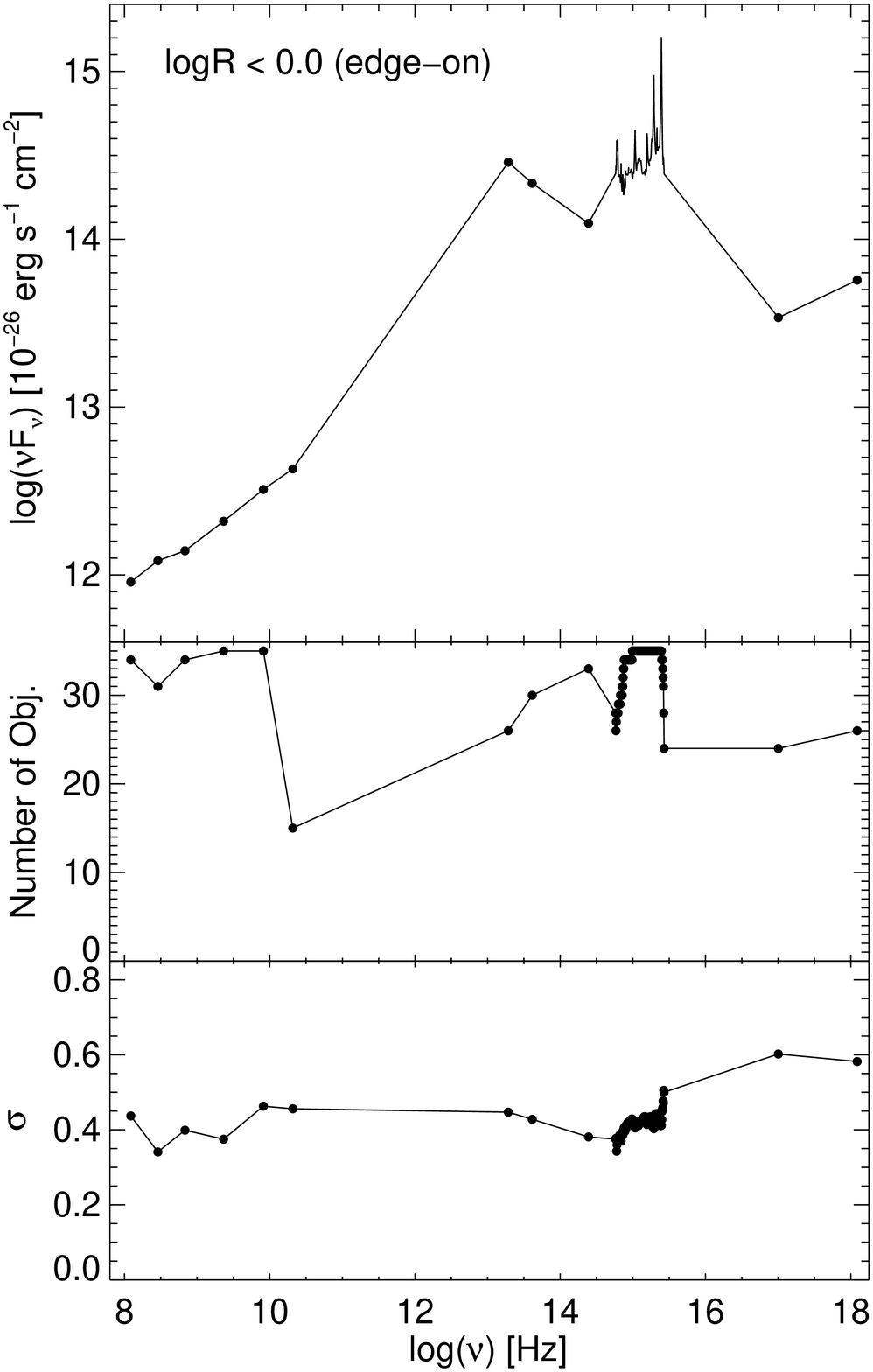}
\end{minipage}
\begin{minipage}[!b]{8cm}
\centering
\includegraphics[width=8.9cm]{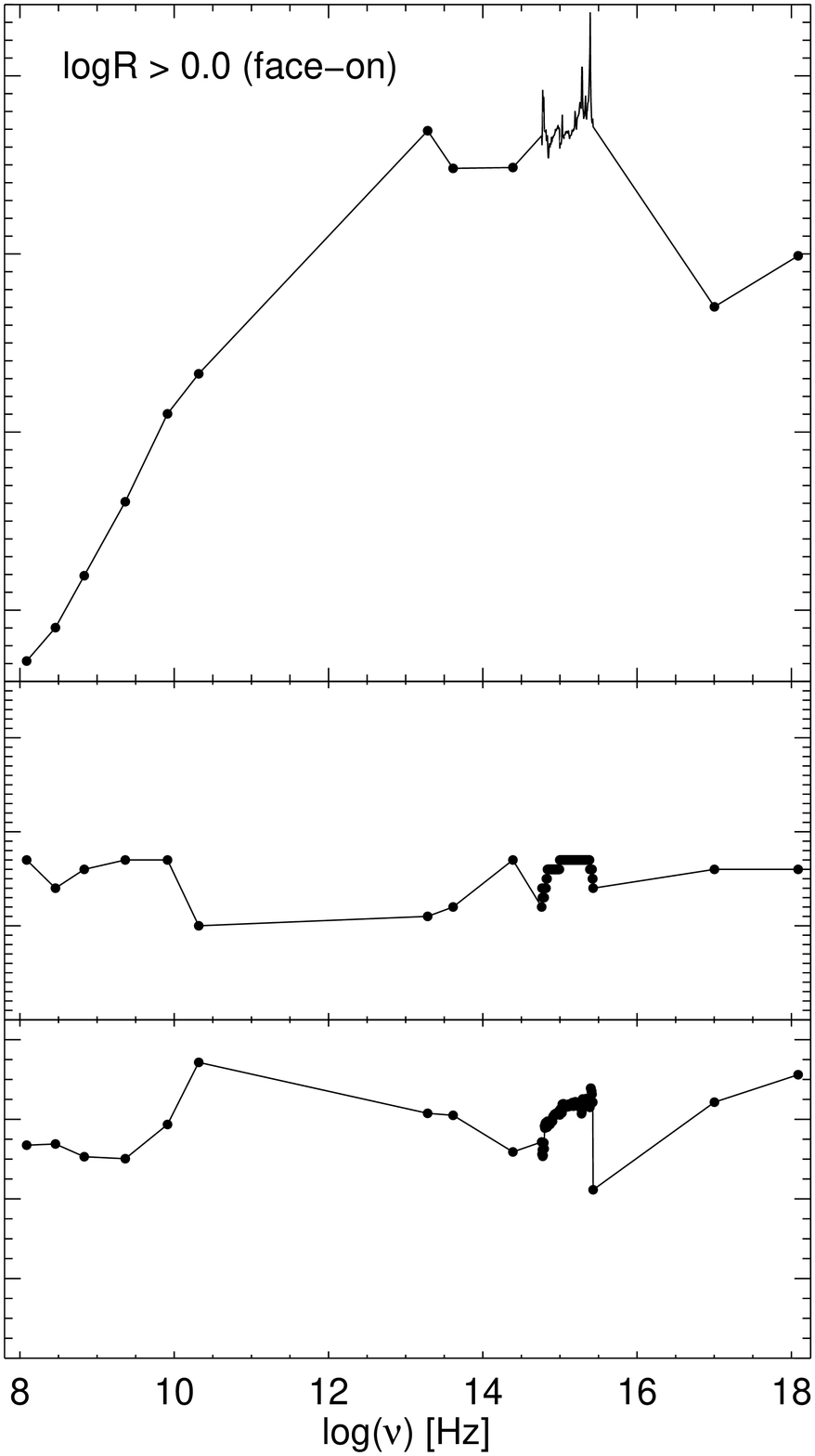}
\end{minipage}
\caption{Composite spectra binned by log~$R$, a radio orientation indicator.  For each composite we also include a panel with the number of sources contributing data to each bin and the standard deviation of the data in each bin.}
\label{fig:composite}
\end{figure*}

To create the final composite, sources with multiple points in one bin are rebinned and then the data are median combined.  This method is chosen so that each source is weighted equally and to reject outliers and prevent individual objects from dominating features of the final composite.  In a final pass, we evaluate the number of objects contributing data in each bin and exclude any points that are determined by fewer than 8 sources.  The final composite SEDs are shown in Figure~\ref{fig:composite} along with panels showing the number of sources and standard deviation in each bin.

In the absence of complete data coverage there is no perfect way to construct composite SEDs.  The partial data coverage in the IR in particular means that those points in the final composite have larger uncertainties associated with them.  However, the composites are intended to be help the reader visualize the results of the correlation analysis so it is not critical to know each point with high precision.  

\begin{figure*}
\begin{center}
\includegraphics[width=17.5 truecm]{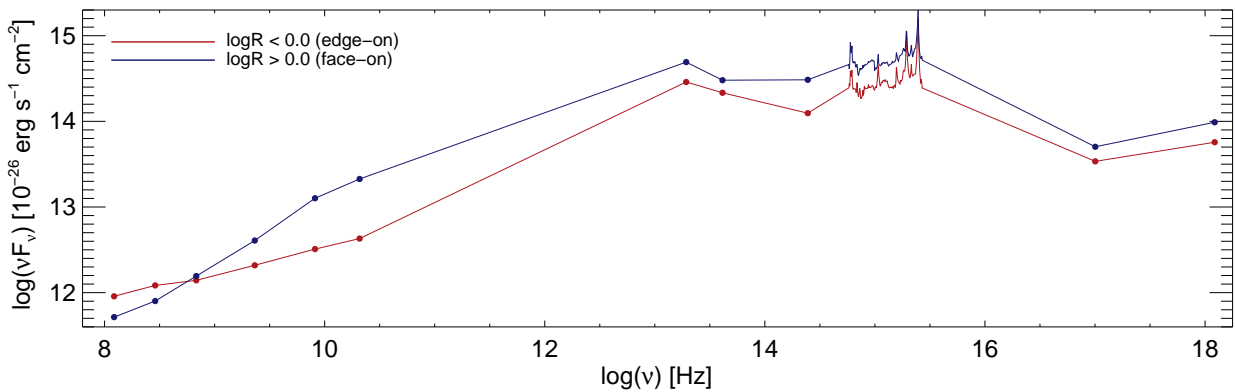}
\end{center}
\caption{Composite SEDs compared over the full frequency range.  The face-on composite is brighter and, most notably, the shape of the SED is similar in all three composites.}
\label{fig:compfull}
\end{figure*}

\begin{figure*}
\begin{center}
\includegraphics[width=17.5 truecm]{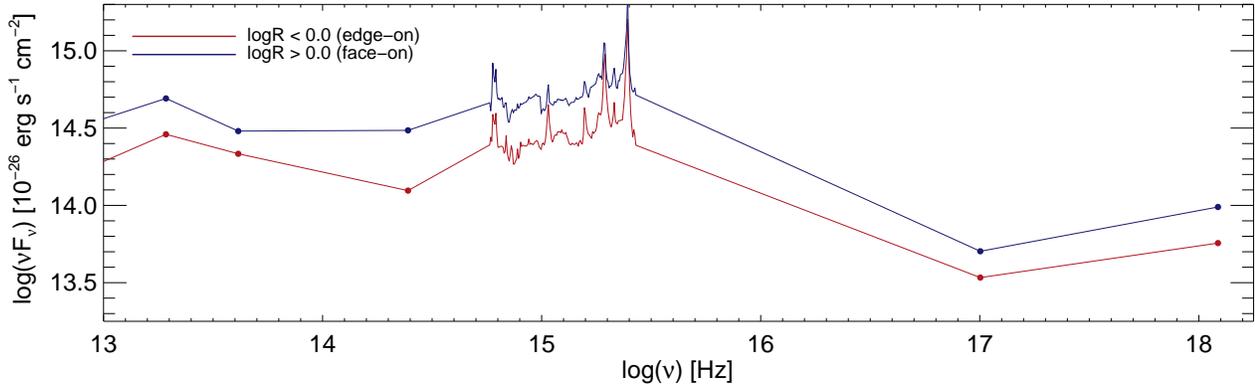}
\end{center}
\caption{Composite SEDs compared in the IR-to-X-ray regime.  The face-on composite is brighter and there is no obvious decrease in the difference between the face-on and edge-on composites with increasing wavelength.}
\label{fig:compzoom}
\end{figure*}

The composites serve to illustrate many of the results found in the rank correlation analysis.  Figure~\ref{fig:compfull} shows the two composites plotted together over the full frequency range.  In general, the face-on composite is brighter than the edge-on composite and the optical-to-X-ray and optical-to-IR slopes are similar between both composites.  It is particularly clear here that the shape of the SED is relatively constant with orientation.  Figure~\ref{fig:compzoom} shows a comparison of the composites in the IR-through-X-ray spectral regime.  In the IR, there is no clear change in the orientation dependence going to lower frequency.

\section{Discussion}
\label{sec:discussion}
The orientation dependence of quasar emission has long been of interest.  As a result, many of these results are known in the literature and we discuss them here in the context of previous work.
	
\subsection{X-ray and optical/UV luminosities}	
The orientation dependence of the X-ray monochromatic luminosity at 2~keV is known \citep{jackson89} and has an accepted explanation.  RL quasars are known to have more X-ray flux than RQ quasars, typically an excess of a factor of 3 but potentially a factor of 10 or more for the most RL sources \citep{miller11}.  This X-ray emission has two components from different origins.  One component, that is likely similar to what is observed in RQ quasars, is consistent with an origin related to a hot corona and the accretion disk \citep{miller11}.  There is also a jet-linked component \citep[e.g., ][]{kembhavi86,shastri93,miller11} that is stronger in objects viewed more face-on \citep{worrall87} and causes objects that are more RL to have flatter X-ray spectra \citep{wilkes87,shastri91}.  As our sample is exclusively RL, it is possible that at least some of the observed orientation dependence is the result of a beamed X-ray component.  Without the ability to separate the X-ray emission from a hot corona and a jet-linked mechanism, it is not possible to conclusively attribute the observed orientation dependence to either source.  In particular, this means that for the hot corona, we can't address the degree of anisotropy.

The orientation dependence of the optical/UV luminosities is known and has been demonstrated for this sample before by \citet{runnoe13a}.  \citet{browne85} and others have shown that core-dominated sources (viewed more face-on, log~$R>0$) are optically brighter than lobe-dominated sources.  Furthermore, the degree of anisotropy observed in this sample is generally consistent with the amount, typically a few tenths of a decade between edge-on and face-on sources, expected from accretion disk models \citep{laor89,nb10}.
		
\subsection{IR luminosities}
Observations are generally consistent with the level of orientation dependence expected from a clumpy torus as opposed to a smooth one (which would have a very strong degree of anisotropy of an order of magnitude or more), but there is variation in the details.  In general, the approach to evaluate the degree of anisotropy in the MIR emission has been to compare the MIR emission normalized to the intrinsic AGN power (usually an X-ray or radio luminosity) in Type 1 and Type 2 sources, which are thought to be viewed face and edge-on, respectively, in the unified model.  

Based on this type of analysis, some investigations conclude that the MIR emission from AGN is relatively isotropic.  Comparisons between Type 1 and Type 2 Seyferts reveals no difference in the relation between the MIR and X-ray luminosities, indicating that the MIR emission is similar in face-on and edge-on objects and therefore is relatively isotropically emitted \citep{horst08,levenson09}.  However, in the case of \citet{levenson09}, there are significant uncertainties on the luminosities used to draw this conclusion.  

On the other hand, some studies find that the MIR is emitted anisotropically.  A comparison of the IR SEDs of RL quasars and radio galaxies reveals a moderate difference where RL quasars are typically a factor of 1.4 brighter at 15~$\mu$m \citep{hoenig11}, however the unification of RL quasars and radio galaxies is questionable (\citealt{singal13a}; DiPompeo et al. in prep.).  Comparisons of the MIR-to-radio flux ratio in Type 1 and 2 Seyferts indicate that the Type 1 sources are brighter at 10~$\mu$m by factors ranging from 4 \citep{heckman95} to $>6$ \citep{buchanan06}.  \citet{maiolino95} finds that the absolute 10~$\mu$m flux is brighter in Type 1 Seyferts.  In a sample of RL quasars with known orientations from radio observations, \citet{landt10} find that the emission from warm and cool dust increases relative to the emission from warm dust, which is inconsistent with smooth torus models.

Our results are broadly consistent with the investigations that find some degree of anisotropy in the MIR emission, although with only Type 1 sources none of our inclinations are truly edge-on.  In actuality, both groups of results (those that find isotropic and anisotropic MIR emission) are consistent with clumpy dusty torus models like those of \citet{nenkova08a}.  The input parameters of these models are degenerate, with several able to control the anisotropy of the MIR emission, and the models are capable of producing significant variation in the degree of anisotropy.  Using a wide range of reasonable input parameters, \citet{levenson09} generate on the order of $10^3$ spectra from the models of \citet{nenkova08a}.  The greatest degree of anisotropy between any two of these spectra is a factor of 600 difference at 8~$\mu$m.  More typically for a single set of input parameters, the difference is less than a factor of 20 between edge-on and face-on spectra.  This is very consistent with the observational results, including ours.  We avoid actually fitting the models to our data because with only 4 data points it is difficult to break the degeneracy between the input parameters that control the anisotropy.  It is possible to distinguish between different realizations, but the preferred method involves the 10~$\mu$m feature and therefore detailed spectra \citep{levenson09}, which is beyond the capabilities of our dataset.

Radiative transfer modeling also predicts trends in the orientation dependence of the IR emission with wavelength.  The models of \citet{nenkova08a} show a decreasing degree of anisotropy with increasing wavelength, whereas we observe no trend in our sample.  The discrepancy may be due to a combination of effects.  First, in the models the anisotropy of the IR emission does not start to significantly decrease until wavelengths longer than are covered in our sample.  Second, because our sample is composed entirely of Type 1 objects, we are unable to probe truly edge-on orientations.  Furthermore, these particular models have been shown to underpredict the FIR flux \citep{landt10}.  So assuming that the models correctly predict the orientation trends in the FIR, while they in general show decreasing anisotropy with increasing wavelengths, at the wavelengths and viewing angles that we probe the decrease is much less significant.  Some observational studies also find a decrease in anisotropy with increasing wavelength \citep[e.g.,][]{buchanan06,hoenig11}, but both studies include Type 2 sources and therefore have a greater range in viewing angles.

\subsection{Covering fraction}	
The covering fraction estimate that we use here assumes that the optical/UV is emitted with a larger degree of anisotropy than the IR.  We find that the orientation dependence in both wavelength regimes is similar, although technically larger in the optical/UV.  While the method of \citet{calderone12} is still valid, it could be simplified with this new information.  By assuming that the degree of anisotropy is similar and following their derivation, the covering fraction can be approximated as $c\approx~\sqrt{R}$.

There are several issues regarding typical calculations of covering fractions that make it likely that estimated values differ from the true quantity.  The covering fraction is estimated by assuming that the dusty torus reprocesses the optical/UV accretion disk emission into the IR.  However, in many cases it is assumed that the disk emits isotropically, which is not the case.  \citet{calderone12}, whose prescription we follow, assume that the disk emits anisotropically, although the description of the orientation dependence is simplistic.  \citet{ma13} simulate the effect of anisotropic disk emission due to limb darkening in a non-relativistic disk.  Emission from the accretion disk is subject to general relativistic effects that also modify the anisotropy of the optical/UV emission \citep[e.g.,][]{fukue06}.  Relativistic effects will bend the light such that the edge-on flux is increased compared to non-relativistic disk.  Thus a $\cos (\theta)$ description of the anisotropy of the accretion disk under-predicts the edge-on flux and over-predicts the covering fraction of the dusty torus.  For this reason, our covering fraction estimates can be thought of as upper limits on the true value.  A more careful calculation of covering fraction would account for these issues, potentially using the corrections available from \citet{nb10}.  There is large scatter associated with converting radio orientation indicators to precise viewing angle measurements, which would make this analysis very uncertain.  	

The potential of contaminating emission is another consideration in covering fraction calculations.  In luminous sources, as in this sample, the AGN dominates at most wavelengths, but contamination can occur.  The host galaxy can contribute a significant fraction of the emission in the far-infrared \citep[FIR,][]{netzer07} and in the NIR and optical.  It is also possible, particularly in RL sources, to have a significant contribution to the IR from synchrotron emission originating in the jet.  In some investigations it is possible to decompose the IR emission to some extent to give a handle on the primary emission source \citep[e.g.,][]{levenson09,landt10}.  

In this sample we have taken some steps in this direction.  \citet{shang11} have corrected the SEDs for contamination from the host in the NIR and optical.  We further avoid these issues by using UV and MIR luminosities to calculate covering fraction, thus avoiding the potential for contamination in derived parameters.  In the FIR, the host is not expected to contribute a significant fraction of the emission until much longer wavelengths than are covered by WISE in this sample.  Finally, the blazars, which can have significant IR variability due to a synchrotron component, have been excluded from this sample.

Given the technical issues associated with estimating covering fraction, it is not surprising that there are quite a range of values in the literature.  We find a larger range of covering fractions than \citet{calderone12} ($0.54-0.70$ compared to our $0.38-0.95$), but there are differences in our method of estimating the integrated IR luminosity that may play a role here.  Following a very similar procedure, \citet{dipompeo13a} finds covering fractions significantly smaller than ours ($0.36$ compared to our $0.56$ on average), but this is largely the result of the fact that they use a monochromatic IR luminosity rather than an integrated one.  We find similar ratios of $L$(MIR)/$L_{bol}$ to \citet{ma13} and typically smaller values than \citet{maiolino07}, indicative of our finding similar and smaller covering fraction estimates, respectively.  The known luminosity dependence of the covering fraction is another issue when comparing covering fraction estimates \citep[e.g.,][]{maiolino07,ma13}; samples must have similar luminosity ranges in order to make a meaningful comparison.  With this in mind we would expect, for the same procedure, to find larger covering fractions than \citet{dipompeo13a}, who have a higher luminosity sample.

The orientation dependence of covering fraction estimates and similarly the optical-to-IR spectral index are of particular interest in the context of radio-quiet orientation indicators.  In general, the radio-loud orientation indicators, for example radio core dominance, function by normalizing an anisotropically emitted quantity to one that is known to be isotropic.  Given that the optical/UV emission is anisotropic, the ability of the \citet{nenkova08a} models to produce isotropic IR emission was potentially of interest.  However, the covering fraction estimates and the IR-to-optical spectral index correlations both indicate that the covering fraction estimates cannot be used as a RQ orientation indicator.

\subsection{Composite SEDs}
As a result of the difficulty of observing the full SED of a sample of objects with radio orientation indicators, composite SEDs as a function of orientation are rare in the literature and often lack complete wavelength coverage.  One common approach to asses the orientation dependence of the SED without a radio orientation indicator is to compare RL quasars and radio galaxies.  Following this method, \citet{haas08} and \citet{hoenig11} both find that quasars are brighter than radio galaxies for SEDs that cover the IR and radio-through-optical, respectively.  This is generally consistent with our composites, which are also brighter for more face-on sources.  The caveat here is again that the unification of radio galaxies and quasars seems more complicated than a simple orientation picture can explain (e.g., \citealt{singal13a}; DiPompeo et al. in prep.).


\section{Conclusions}
\label{sec:conclusion}
We investigate the orientation dependence of a variety of properties of quasar SEDs, including X-ray, UV, and IR monochromatic luminosities, an integrated MIR luminosity, spectral slopes, and the covering fraction of the dusty torus.  We use the RL subsample of the \citet{shang11} SED atlas, ideal for orientation studies, where objects are selected to be intrinsically similar but viewed at different angles to isolate the effects of orientation on the SED.  We perform a rank correlation analysis and create composite spectra to find the following results:

\begin{itemize}

\item  We find statistically significant orientation dependencies in the monochromatic luminosities at 2~keV, 2500, and 5100 \AA\  and marginally significant orientation dependencies in the UV and IR luminosities at 1450~\AA\ and 12, 8, and 3~$\mu$m.  Face-on sources are brighter than edge-on sources in this Type 1 sample, typically by a factor of a 2 in the IR, 2.5 in the optical/UV, and 3 in the X-rays.  The anisotropy at each wavelength is similar so that the shape of the SED does not change significantly with orientation.

\item In the infrared, there is no significant change in the orientation dependence with wavelength in our sample.  Some models predict a decrease in anisotropy with increasing wavelength that is observed in some samples, but it is possible that we are unable to detect this effect given the wavelength range and viewing angles in our sample. 

\item We do not find a statistically significant orientation dependence for the X-ray-to-optical and optical-to-IR spectral slopes, $\alpha_{ox}$ and $\alpha_{oIR}$, respectively.

\item The covering fraction of the dusty torus, estimated from the observed ratio $L(\textrm{MIR})/L_{bol}$ and assuming that the accretion disk emits anisotropically via a geometric $\cos (\theta)$ effect, does not display a statistically significant orientation dependence.

\item Composite SEDs determined for log~$R<0$ and log~$R>~0$ illustrate the conclusions drawn from the correlation analysis, with face-on sources being brighter at all frequencies and no significant orientation dependence in the shape of the SED.

\end{itemize}

\section*{Acknowledgments}
JCR would like to thank Sabrina Cales and Mike DiPompeo for helpful discussions during the preparation of this work.  We also thank the anonymous referee for helpful suggestions.

\bibliographystyle{/Users/jrunnoe/Library/texmf/bibtex/bst/mn2e}
\bibliography{./all.053113}
\clearpage

\label{lastpage}
\end{document}